\documentclass[aps,prb,onecolumn,groupedaddress,showpacs,superscriptaddress,amssymb,amsmath,longbibliography,]{revtex4-2}
\usepackage{amsmath,amsthm}
\usepackage{amssymb}
\usepackage[margin=1in]{geometry}
\usepackage{mathtools}
\usepackage{dsfont}
\usepackage{xcolor}
\usepackage{algorithm2e}
\usepackage{graphicx}
\usepackage{subcaption}
\usepackage{hyperref}
\usepackage{cleveref}
\usepackage{nicefrac}

\usepackage{thmtools}
\usepackage{thm-restate}

\usepackage{mathrsfs}
\usepackage{nccmath}
\usepackage{placeins}
\usepackage{physics}
\usepackage{comment}
\usepackage{qcircuit}
\usepackage{ulem}

\newtheorem{theorem}{Theorem}

\newtheorem{corollary}{Corollary}[theorem]
\newtheorem{definition}[theorem]{Definition}

\usepackage{todonotes}

\newcommand{\bea}{\begin{eqnarray}}
\newcommand{\eea}{\end{eqnarray}}

\newcommand{\trotterchan}[2]{{S_{#1}{\parens{ #2 }}}}

\newcommand{\tschan}[2]{\mathcal{T}^{#1}{( #2 )}}

\newcommand{\liouv}{\mathcal{L}}

\newcommand{\indone}[1]{\norm{ #1 }_{1\rightarrow 1}}

\newcommand{\parens}[1]{\left( #1 \right)}

\begin{document}
\title{Quantum Simulation of Lindbladian Dynamics via Repeated Interactions}
\author{Matthew Pocrnic}
\affiliation{Department of Physics, 60 Saint George St., University of Toronto, Toronto, Ontario,  M5S 1A7, Canada}

%\author[3, 4]{Dvira Segal}
\author{Dvira Segal}
\affiliation{Department of Chemistry and Centre for Quantum Information and Quantum Control,
University of Toronto, 80 Saint George St., Toronto, Ontario, M5S 3H6, Canada}
\affiliation{Department of Physics, 60 Saint George St., University of Toronto, Toronto, Ontario,  M5S 1A7, Canada}
%\email{dvira.segal@utoronto.ca}
%\author[2, 4, 7]{Nathan Wiebe}
\author{Nathan Wiebe}
\affiliation{Department of Computer Science, University of Toronto, Toronto ON, Canada}
\affiliation{Pacific Northwest National Laboratory, Richland Wa, USA}
\affiliation{Canadian Institute for Advanced Research, Toronto ON, Canada}

\begin{abstract}
The Lindblad equation generalizes the Schr\"{o}dinger equation to quantum systems that undergo dissipative dynamics. The quantum simulation of Lindbladian dynamics is therefore non-unitary, preventing a naive application of state-of-the-art quantum algorithms. Here, we make use of an approximate correspondence between Lindbladian dynamics and evolution based on Repeated Interaction (RI) CPTP maps to write down a Hamiltonian formulation of the Lindblad dynamics and derive a rigorous error bound on the master equation.  Specifically, we show that the number of interactions needed to simulate the Liouvillian $e^{t\mathcal{L}}$ within error $\epsilon$ scales in a weak coupling limit as $\nu\in O(t^2\|\mathcal{L}\|_{1\rightarrow 1}^2/\epsilon)$. This is significant because the error in the Lindbladian approximation to the dynamics is not explicitly bounded in existing quantum algorithms for open system simulations. We then provide quantum algorithms to simulate RI maps using an iterative Qubitization approach and Trotter-Suzuki formulas and specifically show that for iterative Qubitization the number of operations needed to simulate the dynamics (for a fixed value of $\nu$) scales in a weak coupling limit as $O(\alpha_0 t + \nu \log(1/\epsilon)/\log\log(1/\epsilon))$ where $\alpha_0$ is the coefficient $1$-norm for the system and bath Hamiltonians. This scaling would appear to be optimal if the complexity of $\nu$ is not considered, which underscores the importance of considering the error in the Liouvillian that we reveal in this work.  

\end{abstract}

\maketitle

\section{Introduction} \label{sec:intro}
The simulation of quantum systems is arguably the most natural application of quantum computers \cite{feynman2018simulating}. The simulation problem can be summarized as follows: For a given Hamiltonian $H$ and evolution time $t$, construct a unitary $U$ that approximates the dynamics generated by $H$ up to an error $\epsilon$, with respect to some appropriate distance metric. In the case of time-independent $H$ we wish to construct a quantum algorithm to build $U$, and approximate the solution to the Schr\"{o}dinger equation $\partial_t \ket{\psi} = -iH\ket{\psi}$ such that $\norm{U - e^{-iHt}}\leq \epsilon$.  
This problem has deservedly received much attention, and the field has achieved many remarkable results \cite{lloyd1996universal, low2019hamiltonian, childs2012hamiltonian, Berry_2015, babbush2023exponential, dalzell2023quantum}. 
However, significantly less attention has been given to non-unitary quantum dynamics, or open quantum systems. These are systems that interact and exchange energy with a larger environment. The field of open quantum systems has put forward a variety of methods for studying such systems to varying levels of approximation and accuracy \cite{Thoss20}. %DS 
In particular, the formalism of quantum master equations (QMEs) has been proven useful to describe dissipative quantum dynamics.
A QME is differential equation for the system's density matrix dictating its evolving in time, incorporating the effects of both coherent evolution and dissipative processes influenced by the environment \cite{OQSbook}. % DS
The approach we implement in this work is the Lindblad quantum master equation \cite{OQSbook, Rivas12}. From the perspective of quantum computing, this is perhaps the most natural master equation to implement given that it generates dynamics described by completely positive trace preserving (CPTP) maps. This equation also generates Markovian dynamics, meaning that correlations between the bath and the system do not build up. Physically, this means that the bath dynamics occur at a much faster time scale than the system, so any correlations in the bath due to the coupled system propagate to infinity before they influence the system once again. Conceptually, one would therefore expect to be able to simulate Markovian dynamics in a smaller Hilbert space, and therefore, with less resources than non-Markovian quantum master equations. \\

In this work, we present quantum algorithms based on an existing approximate correspondence between the Lindblad equation and repeated interaction (RI) maps, also known as collision models \cite{ciccarello2022quantum, bruneau2014repeated,strasberg2017quantum, guarnieri2020non, purkayastha2021periodically}. Repeated interactions model is a scheme whereby the environment repeatedly interact with the system. 
The environment is comprised of many independent bath subsystems with each interacting with the system one by one.
These baths/subsystems do not interact with one another, and never interact with the system again, naturally leading to a Markov process. While this might seem contrived, open quantum systems such as the maser can be naturally described in this fashion: In this case,  an electromagnetic field in a cavity is repeatedly perturbed by particles passing through, whereby the reduced dynamics of interest is that of the quantum field \cite{ciccarello2022quantum, bruneau2014repeated}.
Here, we provide quantum algorithms to implement these repeated interactions to effectively model Lindbladian dynamics. Our algorithms yield integrators of arbitrary order, which we expect to be nearly optimal given that the error between the Lindblad dynamics and those generated by repeated interactions is second order in the interaction time in general. Implementing low order formulas here is not as problematic as they may seem given the physics captured by the Lindblad equation: This QME can be derived by integrating the exact equation of motion generated by an interaction Hamiltonian $H_I$ that governs the entire system bath dynamics \cite{OQSbook}. This integral equation is then expanded to second order in the interaction time, and a trace is performed over the environment Hilbert space. This therefore gives an accurate description of the exact dynamics only up to second order, so we might say the Lindblad equation roughly has model error $\epsilon_{model} \in O(||H_I||^3)$. This implies that using higher order integrators to simulate dynamics to incredibly small error $\epsilon$ 
is not necessarily capturing additional physics. Note that the model error is inherently different than the \textit{simulation error} $\epsilon$ which will be the main error of interest in this paper, as it quantifies how our algorithm deviates from Lindblad dynamics. The former \textit{model error} is not analyzed in this work. \\

\section{Contributions and Related Work}
When it comes to related works, state of the art Lindbladian simulation is presented in Refs.\cite{cleve2016efficient, li2022simulating}. The work of Cleve and Wang Ref.\cite{cleve2016efficient}, utilizes Kraus maps to approximate the Lindbladian to second order, and then constructs a variant of linear combinations of unitaries (LCU) for channels to implement the Kraus maps in a purified space. The approach then utilizes a compression encoding scheme involving the Chernoff bound to achieve complexity $O(t\text{polylog}(t/\epsilon))$. In addition, Ref.\cite{cleve2016efficient} also provides an early discussion of simulating a Lindblad equation with iterative unitary evolutions, however, this approach differs from the RI scheme we present here. Their work briefly examines implementing a block matrix $J$ built out of all system jump operators $L_j$, and iteratively applying $e^{-iJ\sqrt{\tau}}$ and $e^{-iH\tau}$ with trace operations in an alternating fashion, rather than iteratively interacting with subsystems in the RI fashion we study here. Regarding this approach, in the appendix of their work, a lower-bound is given with regards to expressing Lindbladian evolution as effective Hamiltonian evolutions. The Theorem concludes that in general, a Hamiltonian evolution cannot \textit{exactly} yield Lindblad dynamics for any \textit{finite} time. While we do not necessarily disagree with this statement, it does not prevent choosing an iteration parameter to make $\epsilon$ arbitrarily small. This bound is also derived such that their algorithm approximates a joint system-bath evolution with trace-out as a purely unitary process on the system with an effective Hamiltonian. Given that our algorithms do not inherit this strategy, this lower bound does not apply. In addition, in \cite{cleve2016efficient}, Cleve and Wang also use this early iterative Lindblad simulation to suggest that expressing said dynamics with Hamiltonian evolutions should incur overhead that scales like $O(t^2/\epsilon )$. We make this intuition rigorous, and arrive at a similar conclusion in Theorem \ref{thm:RI_error}. Next, the approach in Ref. \cite{li2022simulating} 
utilizes an interesting application of Duhamel's principle to build high-order integral formulas, which are approximated as a sum, derived from Gaussian Quadrature methods. Linear combinations of unitaries are then constructed once again for implementation, again achieving time complexity $O(t\text{polylog}(t/\epsilon))$, with slightly better dependence on the number of dissipators in the Lindbladian. While Refs.\cite{cleve2016efficient, li2022simulating} achieve state-of-the-art asymptotics, they both involve intricate machinery that make it unclear as to how well they will perform in practice, the former relying on a compression encoding scheme, and the latter involving the computation of multi-integrals of block encoded Kraus operators using LCU circuits. A goal of our work is to build conceptually cleaner algorithms that, in comparison, still perform well asymptotically. Other earlier notable algorithmic contributions in this field can be found in Refs.\cite{2017, barthel2012quasilocality, kliesch2011dissipative}. Alternative implementations involving unitary decompositions that also conduct numerical experiments on simulators can be found in Refs. \cite{schlimgen2021quantum, schlimgen2022quantum}. 
Most recently \cite{ding2023simulating}, work has been done to formulate integrators of arbitrary order based on repeated Hamiltonian simulations, which is reminiscent of the theme of this work. 
The authors in Ref.\cite{ding2023simulating} propose a strategy that is also based on maps of a Stinespring form with $\sqrt{\tau}$ evolution of an effective Hamiltonian. However, this approach is derived from a method known as \textit{unravelling}, in which the Lindblad equation is transformed into a stochastic Schr\"{o}dinger equation (SSE). Interestingly, the $\sqrt{\tau}$ Hamiltonian evolution $e^{-iH\sqrt{\tau}}$ that arises from the variance of the Wiener process in their work also arises in the RI maps in our work for seemingly unrelated reasons. Equipped with the SSE, the goal is to solve the dynamics for the expectation value of the final state. This approach is similar to an earlier correspondence proposed between the repeated interactions scheme that we investigate here, and quantum Langevin equations \cite{attal2006repeated}. The intuition between this relation is that coupling of the system to a random ancilla can lead to the generation of random \textit{quantum forces}. Ref. \cite{ding2023simulating} also generalizes results to time dependent Lindbladians, a topic that has seldom been studied in the context of quantum simulations. Recent work has also used a product formula decomposition of the Lindblad evolution into Hamiltonian and dissipative parts, and then implemented a sampling routine to simulate the latter part \cite{borras2024quantum}. Lastly, complimenting the many above recent studies on upper-bounding the complexity of simulating open Markovian dynamics, Ref. \cite{ding2024lower} provides novel circuit complexity lower bounds. \\

Regarding the method of formulating Linadbladian dynamics as repeated Hamiltonian simulations, other recent works have also emerged \cite{di2023efficient, patel2023wave, patel2023wave2}. Ref \cite{di2023efficient} can be viewed somewhat as a hybrid approach between that of Ref. \cite{ding2023simulating} and our own. In Ref. \cite{di2023efficient} the authors implement the solution to an \textit{unravelled} Lindblad equation via stochastic unitary processes that are implemented via Trotterization and repeated trace-out. The given complexity of $\epsilon$ is also quadratic in time with additional overheads that arise from making locality assumptions required to discretize infinite dimensional bath operators.  In addition, a somewhat different approach called Wave Matrix Lindbladinization has also been proposed \cite{patel2023wave, patel2023wave2}. Similar to our work, this method involves the formulation of a map that approximates a Lindbladian evolution to second order in time. It is then implemented in an iterative fashion with trace-out to control $\epsilon$. However, the input model is quite different from other works in that the algorithm assumes access to copies of so-called \textit{program states}, which are quantum states that encode the Lindbladian itself. Bounds are then given in terms of queries to an oracle that provides said states, which interestingly inherits similar time-complexity to the parameter $\nu$ discussed in this work. 
\\ 

In terms of our contributions, we were largely inspired by Ref. \cite{cattaneo2021collision}, which to the best of our knowledge, was the first to propose quantum simulation of the RI scheme. 
Our work builds upon yet differs significantly from this analysis. Here, the implementation of the RI maps is abstracted from the physical picture, whereby we analyze the complexity of converging to nearly arbitrary time-independent Lindbladian dynamics using an RI map with minimal assumptions. In doing so, we provide a rigorous error bound 
on the difference 
between these two CPTP maps. This is one of the main contributions of our work, which to the best of our knowledge has not been done before in this context. Bounds given in Ref. \cite{cattaneo2021collision} solve a problem with a physical model in mind. There, a k-local Lindbladian is assumed to be simulated with a physical multipartite interaction model in which subsystems interact locally with a system governed by $H_S$, supported on some lattice. This necessarily involves some complicated indices such as the set of all ancilla-system site pairings, which can make the bounds difficult to interpret. As well, for the purpose of simulation, the analysis in Ref. \cite{cattaneo2021collision} uses Trotter-Suzuki formulas to implement each propagator from the beginning of the analysis. Here, we specifically bound the convergence of an RI map to Lindbladian dynamics, where we use the number of interactions with ancillary subsystems to control the error. We then use modern quantum algorithms based on block encodings to implement these general maps as efficiently as possible. For means of simplicity and comparison, we further provide a simulation of our RI maps with Trotter-Suzuki formulas. We also note that multipartite and local dynamics can be simulated with our model by restricting the support of the operators in the unitary decomposition of the interaction Hamiltonian. We summarize our contributions below: 

\begin{enumerate}
    \item We rigorously bound the error between RI maps and Lindbladian dynamics in terms of the number of environment subsystems $m$ and the number of applications $\nu$ of an RI map composed of $m$ interactions in Theorem \ref{thm:RI_error}.
    \item In Corollary \ref{cor:kappa}, we then immediately obtain a bound on the number of ancilla bath's subsystems, given that the Trace is performed over the entire environment register at the {\it end} of the simulation.
    \item We describe a strategy to achieve the same \textit{simulation accuracy} and time complexity using a constant number of qubits, given that traces are performed periodically.
    \item For either of the ancilla strategy above, we implement a simulation of Lindbladian dynamics using unitary RI maps, with a probability of success $\delta=1$ and an error $\epsilon$, utilizing an iterative Qubitization strategy to implement each discrete time map. 
    We give a bound on the time complexity in terms of queries to block encoding oracles in Theorem \ref{thm:qubitization}.
    \item To compare with Ref. \cite{cattaneo2021collision}, we implement our RI maps with Trotter-Suzuki formulas, and bound the number of steps/iterations $r$ 
    to achieve error $\epsilon$ for formulas of first and arbitrary order, the former of which we utilize commutator structure. 
    This is presented in Theorems \ref{thm:Trotter2k} and \ref{thm:first_trotter}.
\end{enumerate}

\section{Preliminaries}
In this Section, we introduce the relevant notations and assumptions for the work. We also provide a brief introduction to the Lindblad equation and repeated interaction  maps, and some of their relevant properties. 

\subsection{Notation and Assumptions}
We use lowercase letters to denote variables, $a$, uppercase letters to represent operators, $A$, and reserve calligraphic letters for super-operators, $\mathcal{A}$. In this paper, we consider only finite-dimensional Hilbert spaces, so all operators $A$ are matrices, and all super-operators can be written as some functions of matrices. An important part of analyzing quantum algorithms is bounding the error $\epsilon$, which allows one to calculate the complexity of the algorithm. Since the algorithms here are written in terms of super-operators, we implement an induced norm to calculate $\epsilon$. The familiar Schatten p-norms for matrices are defined in the following way: $\norm{A}_p = (\sum_k s_k^p )^\frac{1}{p}$, where $s_k$ represent the singular values of the matrix $A$. The Schatten norms then induce a norm on super-operators, that of focus here will be the induced $1\rightarrow 1$ norm, which has the following definition: $\indone{\mathcal{A}} = \max_{A : \norm{A}_1 = 1}\norm{\mathcal{A}(A)}_1$. This norm obeys the triangle inequality and submultiplicativity, which we will make use of. Throughout the work, we write $\rho$ to indicate a density matrix, which is a positive semi-definite operator with $\Tr \rho =1$. Some other recurring notations are the use of $\liouv$ as a Lindblad generator, $t$ for the total simulation time, $m$ as the number of subsystems or ancilla used in the repeated interactions scheme, $\nu$ as the number of applications of an RI map with $m$ interactions, and $\tau = t/\nu$ as the interaction time interval. We use $\circ$ to indicate the composition of super-operators, all of which are quantum channels in this work. When the composition is indexed over many channels, we use $\bigcirc$ to be distinct from a product.

\subsection{Lindblad Equation}
The Lindblad equation, often seen as a generalization of the Schr\"{o}dinger equation for Markovian open quantum systems, takes the following form $(\hbar\equiv1)$:

\begin{equation} \label{eq:lindblad}
    \frac{d \rho_S}{dt} = -i[H_0, \rho_S] + \sum_{j=1} ^m L_j \rho_S L_j^\dagger - \frac{1}{2} \{L_j^\dagger L_j , \rho_S \} \equiv \liouv(\rho_S), 
\end{equation}
 where the $L_j$ operators are derived from the interaction between the system and environment. Therefore, this equation describes the dynamics of the reduced state $\rho_S$ of a quantum system, and does not track the dynamics of the environment. If $L_j = 0 ~ \forall ~ j$ then we recover the von Neumann equation, or the Sch\"{o}dinger equation for density matrices. Here, the Hamiltonian is a Hermitian operator, however, the jump operators $L_j$ are not necessarily so. We also have $H, ~ L_j \in \mathbb{C}^{d \times d}$, where $d$ is the dimension of the system.

\subsection{Repeated Interactions Model}
Repeated interactions allow us to study the reduced dynamics of a quantum system under the influence of an environment, essentially through repeated applications of Stinespring's dilation theorem. The conceptual picture of the RI scheme is that of a quantum system which, for short intervals, repeatedly interacts with an environment comprised of small discrete subsystems or ancilla, which are disjoint. This means that the subsystems comprising the environment do not interact with each other. In addition, once interacting with the quantum system of interest, the subsystems travel off to infinity and never interact with the system again. It is intuitive that such a construction leads to a Markov process, and it turns out that under certain conditions on the coupling energies, this scheme approximates an evolution generated by the Lindblad QME discussed in the previous Section \cite{bruneau2014repeated, ciccarello2022quantum}. Studies regarding a more general RI form including interacting ancilla and recurring interactions have also been studied, resulting in non-Markovian dynamics, as expected \cite{pellegrini2009non}. However, this generalization is beyond the scope of our work. We now present the formalism for the Markovian repeated interactions scheme described above. The system Hamiltonian $H_0$ lives in the Hilbert space $\mathscr{H}_0$, and the $n$th subsystem of the environment lives in the smaller Hilbert space $\mathscr{E}_n$. We represent the entirety of the environment space as the tensor product of all said subsystems:
\begin{equation}
    \mathscr{H}_E = \bigotimes_{n=1}^m \mathscr{E}_n, 
\end{equation}
and the total Hilbert space of the problem is
\begin{equation}
    \mathscr{H} = \mathscr{H}_0 \otimes \mathscr{H}_E.
\end{equation}
We write the total Hamiltonian $H$ in the following way,
\begin{align}
    H = H_0 + \sum_{n=1}^m (H_{E_n} + \lambda H_{I_n}).
\end{align}
The operators act as identities outside the support of their respected subspace, such that it is implied that $H_0 \equiv H_0 \otimes \openone_{\mathscr{H}_E}$, and so on. 
The interaction Hamiltonian takes a tensor product form: $H_{I_n} = V^\dagger \otimes a_n + h.c.$, which is required for the repeated interaction formalism; $\lambda$ is the system-bath coupling parameter. Here, $V$ is some general system operator, and $a_n$ is  an annihilation operator on the $n$th subsystem. For example, $a_n$ could be a bosonic annihilation operator or the spin lowering operator. This total Hamiltonian $H$ generates the repeated interactions map $\mathcal{R}_n$ according to the following definition:

\begin{definition}[Repeated Interactions Map] \label{def:RI}
    Given a system density operator $\rho_S$ of $\dim =d$ that evolves under the free Hamiltonian $H_0$, and a discrete environment comprised of $m$ subsystems $\rho_{E_n}$ of $\dim=2$ that evolve freely under $H_{E_n}$ and interact with $\rho_S$ via an interaction Hamiltonian $H_{I_n}$ with coupling  $\lambda$, then the finite time reduced dynamics of the system $\rho_S(t_0) \rightarrow \rho_S(t_0+\tau)$ is generated by the following repeated interactions map $\mathcal{R}_n$:

    \begin{equation}
    \mathcal{R}_n(\rho_S) = \Tr_{\mathscr{E}_n} \parens{e^{-i\tau (H_0 + H_{E_n} + \lambda H_{I_n})} \rho_S \otimes \rho_{E_n} e^{i\tau (H_0 + H_{E_n} + \lambda H_{I_n})}},
\end{equation}
with interaction time $\tau$. This map can be iterated to produce the system dynamics for long times.
\end{definition}

 Note that the map here is applied only to $\rho_S$, and part of the action of the map is to tensor on the subsystem $\rho_{E_n}$. The repeated interactions map is necessarily applied to the joint system-environment space, but due to the trace it yields the reduced dynamics of the system as required. These maps are Markovian by construction, and given the simplest case where each of the $n$ bath/environment subsystems are identical, we also have the semi-group property:
\begin{equation}
    \mathcal{R}^{\circ m} = \mathcal{R}^{\circ (m-q)} \circ \mathcal{R}^{\circ q},
\end{equation}
for $q < m$, which shows that any composition of RI maps is also an RI map. The semi-group property also highlights the Markovianity of the map: The map can be used to generate any future evolution of the system density matrix from any previous state without access to any notion of the state's history. This is always true for RI maps of this form, given that the map consists of the operations of tensoring on and tracing out the ancillary system. If we considered a hypothetical RI map that acted on a system and ancilla for finite time $\tau$ without the tensor operation, then the Markov property on the joint system only holds in the limit $\tau \rightarrow 0$. These maps do not form a complete group as their inverses are not guaranteed to exist \cite{ciccarello2022quantum}. In this work, to maintain generality, we do not assume the bath qubits to be identical, and subscript the bath's subsystems by $n$. Another important property is that the discretization of the dynamics in this way yields no Trotter error. This is because $[H_0, H_{E_n}] = 0$, as is clear from above, and the interaction $H_{I_n}$ is only turned on for the interaction time $\tau$, similar to a delta kick Hamiltonian. To illustrate, one could equivalently write the map as the following time dependent Hamiltonian evolution with indicator functions
\begin{equation*}
    \bigcirc_{n=0}^{m-1} \mathcal{R}_n(\rho_S) = \Tr_{\mathscr{E}} \parens{e^{-i\tau H(t)} \Bigr (\rho_S \bigotimes_{n=0}^{m-1} \rho_{E_n} \Bigr ) e^{i\tau H(t)}},
\end{equation*}
 with Hamiltonian
 \begin{equation*}
    H(t) = H_0 + \sum_{n=0}^{m-1} \Bigr (H_{E_n} + \lambda H_{I_n} \Bigr ) \Bigr ( \theta(t-n\tau ) - \theta(t-(n+1)\tau \Bigr ),  
\end{equation*}
  where for this example, $\theta(x)$ is the Heaviside function (of which the difference of two serves as a square-pulse or top-hat function), and total evolution time $t = \tau m$. In this form, the absence of a Trotter error is clear. Writing the map for a full evolution in this form does not provide more physical insight, nor does it make for simpler analysis, so we will result to the form of Definition \ref{def:RI} in the remainder of the work. For more properties and applications of RI maps, consult the following reviews \cite{ciccarello2022quantum, bruneau2014repeated}. The most important feature discussed in both, is the correspondence between RI maps and Lindblad dynamics. Below, we give a similar derivation to what can be found in \cite{ciccarello2022quantum, bruneau2014repeated}. However, we provide a more rigorous treatment than what is done in Refs. \cite{ciccarello2022quantum, cattaneo2021collision}, and show a simpler derivation than that contained in Ref. \cite{bruneau2014repeated}.

\section{Lindbladian Derivation} \label{sec:lindblad_derivation}

Let us now show that in a specified limit, the repeated interactions scheme converges to Lindbladian dynamics in the limit of infinitesimally small interaction time. This is provided more so as a demonstration and is not a novel result. Similar work can be found in Refs. \cite{bruneau2014repeated, ciccarello2022quantum, cattaneo2021collision}. However, our derivation tries to achieve the goal of being both pedagogical and rigorous. We will work in the regime where $\lambda^2 \tau =1$, and will study the limit $\tau \rightarrow 0$. Given the nature of this limit, we can think of the repeated interactions as a protocol of strongly coupling the system to the bath's subsystem for very short times. This also provides a somewhat interesting physical picture, given that the Lindblad equation is normally derived assuming weak coupling. 

To derive the Lindblad equation, we write the map in Liouvillian super-operator form and expand for small interaction times,
\begin{align}
    \mathcal{R}_n(\rho_S) = \Tr_{\mathscr{E}_n} \parens{\exp(-i\tau [(H_0 + H_{E_n} + \lambda H_{I_n}), ~ \cdot ~]) (\rho_S \otimes \rho_{E_n})}, 
\end{align}
or we define a super operator $\mathcal{H}(\rho) = -i[(H_0 + H_{E_n} + \lambda H_{I_n}), \rho]$ and equivalently write
\begin{align}
    \mathcal{R}_n(\rho_S) = \Tr_{\mathscr{E}_n} \parens{e^{\tau \mathcal{H}} (\rho_S \otimes \rho_{E_n})}.
\end{align}
For each application, this map effectively performs a dilation by tensoring on an environmental subsystem, followed by a unitary evolution and a reduction. The map is CPTP by Stinespring's dilation theorem. For a single application of the map lets expand to second order in the interaction time

\begin{align}
    \mathcal{R}_n(\rho_S) &= \Tr_{\mathscr{E}_n} \parens{e^{\tau \mathcal{H}} (\rho_S \otimes \rho_{E_n})} 
    \nonumber\\
    &= \Tr_{\mathscr{E}_n} \Biggr ( \rho_S \otimes \rho_{E_n} \notag - i\tau [(H_0 + H_{E_n} + \lambda H_{I_n}), \rho_S \otimes \rho_{E_n}] 
    \nonumber\\ 
    & \quad - \frac{\tau^2}{2} [(H_0 + H_{E_n} + \lambda H_{I_n}),[(H_0 + H_{E_n} + \lambda H_{I_n}), \rho_S \otimes \rho_{E_n}]] \Biggr ) 
    \nonumber\\
    &= \rho_S -i\tau [H_0, \rho_S] - \Tr_{\mathscr{E}_n}\parens{\frac{\tau^2}{2} [(H_0 + H_{E_n} + \lambda H_{I_n}),[(H_0 + H_{E_n} + \lambda H_{I_n}), \rho_S \otimes \rho_{E_n}]]},
    \label{eq:Rn} % DS
\end{align}
In the last step we use the fact that $\Tr_E [H_E, \rho_S \otimes \rho_{E_n}] = 0$. As well, we can either redefine the interaction Hamiltonian, or choose the initial environment state such that $\Tr_E [H_I, \rho_S \otimes \rho_{E_n}] = 0$, as is a standard choice in the derivation of Lindblad master equations.  This leads to
\begin{align}
    \mathcal{R}_n(\rho_S) &= \rho_S - i\tau [H_0, \rho_S] - \frac{\tau^2}{2} \Tr_{\mathscr{E}_n} \Biggl(\lambda^2 \bigl[H_{I_n},\bigl[H_{I_n}, \rho_S \otimes \rho_{E_n}\bigr]\bigr] + \lambda \bigl[(H_0 + H_{E_n}), \bigl[H_{I_n}, \rho_S \otimes \rho_{E_n}\bigr]\bigr] \notag \\
    &\quad + \lambda \bigl[H_{I_n}, \bigl[(H_0 + H_{E_n}), \rho_S \otimes \rho_{E_n}\bigr]\bigr] + \bigl[(H_0 + H_{E_n}), \bigl[(H_0 + H_{E_n}), \rho_S \otimes \rho_{E_n}\bigr]\bigr]\Biggr)
\end{align}
By taking $\lambda \rightarrow \frac{1}{\sqrt{\tau}}$ we get
\begin{align} \label{eq:lambdalim}
    \widetilde{\mathcal{R}_n}(\rho_S) = \rho_S -i\tau [H_0, \rho_S] - \frac{1}{2} \Tr_{\mathscr{E}_n} \Biggl (\tau [H_{I_n},[H_{I_n}, \rho_S \otimes \rho_{E_n}]] + \tau^{3/2} [(H_0 + H_{E_n}), [H_{I_n}, \rho_S \otimes \rho_{E_n}]] \notag \\ + \tau^{3/2} [H_{I_n}, [ (H_0 + H_{E_n}), \rho_S \otimes \rho_{E_n}]] + \tau^2 [(H_0 + H_{E_n}), [(H_0 + H_{E_n}), \rho_S \otimes \rho_{E_n}]] \Biggr), 
\end{align}
where we use a tilde to indicate that this limit has been enacted. This replacement of variables is thoroughly discussed in Ref. \cite{bruneau2014repeated}. 
Rearranging and taking the limit as $\tau \rightarrow 0$ leads to
\begin{align}
    \lim_{\tau \rightarrow 0} \parens{\frac{\widetilde{\mathcal{R}_n}(\rho_S) - \rho_S}{\tau}} &= -i[H_0, \rho_S] - \frac{1}{2} \Tr_{\mathscr{E}_n} \parens{ [H_{I_n},[H_{I_n}, \rho_S \otimes \rho_{E_n}]]} 
    \nonumber\\
    & =  -i[H_0, \rho_S] + \Tr_{\mathscr{E}_n} \parens{H_{I_n} \rho_S \otimes \rho_{E_n} H_{I_n} - \frac{1}{2}\{H_{I_n}H_{I_n}, \rho_S \otimes \rho_{E_n}\}}, 
\end{align}
which is in Lindblad form. 
To then eliminate the trace and write this solely in terms of the system Hamiltonian, we can use the full form of $H_{I_n} = V\otimes a^\dagger + h.c.$. Following \cite{bruneau2014repeated}, if we chose the environment to be in some thermal state at inverse temperature $\beta$, we obtain the following
\begin{align} \label{eq:derived_lindblad}
    \lim_{\tau \rightarrow 0} \parens{\frac{\mathcal{R}_n(\rho_S) - \rho_S}{\tau}} = -i[H_0, \rho_S] + \sum_{j=1}^2\parens{L_j \rho_S L_j^\dagger - \frac{1}{2} \{L_j ^\dagger L_j, \rho_S\}},
\end{align}
where $L_j = V \sqrt{z}$ for $j=1$, $L_j = V^\dagger \sqrt{z-1}$ for $j=2$, and $z = \nicefrac{\Tr_{\mathscr{E}_n} \parens{a_n^\dagger \exp(-\beta a_n^\dagger a_n) a_n}}{ Z(\beta)}$. Here, the $a^\dagger$ and $a$ operators are the creation and annihilation operators, and $Z$ is the partition function. Later we will consider an implementation where the environment is comprised of single qubit thermal states, and as a result, we do not have to worry about exchange statistics. For this reason, $a^\dagger$ and $a$ are sometimes referred to as qubit creation and annihilation operators. With these definitions, the full RI map, the $j$ index will grow to have $2m$ terms for the entire repeated interaction map, where $m$ is the number of baths/environment subsystems or ancilla. For the simple case of a single interaction with the RI map, we have therefore shown how the Lindblad equation can be derived. 

\section{Error Bounding in the Limiting RI case}\label{sec:errorbound}
In this Section we prove an error bound between $\nu$ applications of a RI map composed of $m$ interactions, and Lindbladian dynamics. We work here only in the regime where 
$\lambda = \frac{1}{\sqrt{\tau}}$, using the RI map defined in Equation \ref{def:RI}. 
We can attain equivalent results by using some constant: $\lambda^2 \tau \rightarrow \Gamma$. However, we allow this freedom of the Lindblad jump coefficients to be taken into account through the spectral norm of the Interaction Hamiltonian. 
Working in this limiting case can be regarded as a \textit{replacement of variables} to effectively control the error of the approximation. Working with a general $\lambda$ yields error in the first non-trivial order, which is unfavourable. To see this argument worked out, we refer readers to Appendix \ref{sec:appendixA}. As in the previous Section, we use a tilde to remind the reader that this limiting case of the RI map is being enforced, and write a single application of the map in the following way: $\widetilde{\mathcal{R}_n}(\rho_S)$. This condition is understood to be enforced in the Taylor series expansion of the super-operator, as was also considered in the above derivation. For generality, we leave interactions and the free Hamiltonians of the environment ancilla subscripted with $n$. In this way, the bath can be prepared as an ensemble of single qubit thermal states with different temperatures $\beta$, if the user wishes. Our next objective is to show convergence of the RI map to a general Lindblad QME. 
Given we define the interaction time as $\tau = t/\nu$, where $\nu$ is just some iteration/convergence parameter, the RI map we use to converge to this dynamics, with an arbitrary accuracy, takes the following form
\begin{equation} \label{eq:full_RI}
    \widetilde{\mathcal{R}} = \Bigr (\bigcirc_{n=1}^m \widetilde{\mathcal{R}_n} (\nicefrac{t}{\nu}) \Bigr )^{\circ \nu},
\end{equation}
where $\bigcirc$ indicates an indexed multi-composition
and $\widetilde{\mathcal{R}_n}$ is given by Eq. (\ref{eq:Rn}) upon substitution of $\lambda^2\tau=1$.
Intuitively, an immediate issue arises, in that the general Lindblad picture consists of the system simultaneously coupled to many degrees of freedom, undergoing evolution coupled all subsystems for a time $t$, while the RI map simulates the free evolution for a time $t$, and evolves for a time $\nicefrac{t}{m}$ under each of the $m$ ancilla. To deal with this, we simply re-normalize the free Hamiltonian of the system $H_0 \rightarrow \frac{1}{m}H_0$, which allows us to converge to the following general Lindbladian, 
\begin{equation} \label{eq:rescaled_Lindblad}
    \liouv(\rho_S) = -i[H_0, \rho] - \frac{1}{2}\sum_{k=1}^m\Tr_{\mathscr{H}_E} [H_{I_k}, [H_{I_k}, \rho_S \otimes \rho_{E_k}]] .
\end{equation}
 We can use the trace to extract the more natural jump operators later, as was done in the derivation of Equation (\ref{eq:derived_lindblad}). In Appendix \ref{sec:appendixB}, we provide some sufficiency criteria on $H_{I_n}$ to formulate an RI map that recovers the more familiar Lindbladian in \ref{eq:lindblad}, and in doing so, show how to derive $L_j$. With this in mind, we provide an error bound in the following Theorem:

\begin{theorem}[RI-Lindblad Correspondence Error] \label{thm:RI_error} Given an RI map from Definition \ref{def:RI} and a Lindblad Equation \ref{eq:rescaled_Lindblad}, then the induced Schatten 1-norm of their difference has the following upper bound:
\begin{align}
    \indone{e^{t \liouv} - \Bigr (\bigcirc_{n=1}^m \widetilde{\mathcal{R}_n} (\nicefrac{t}{\nu}) \Bigr )^{\circ \nu}} &\in O \Biggr ( \frac{t^2}{\nu} \Biggr (\indone{\liouv}^2 + m \max_n \gamma_n^4 \Biggr ) \Biggr ),
\end{align}
where $\nu$ is the number of applications of the RI map, $t$ the total simulation time, $m$ the number of subsystems in the bath, and $\gamma_n = \max \{\norm{H_0}, \norm{H_{E_n}}, \norm{H_{I_n}} \}$.
\begin{proof} 

Our goal is to show that in terms of the time scaling we have error that goes like $O(\tau^2)$, in accordance with other analyses in the literature \cite{bruneau2014repeated, cattaneo2021collision}. 

We begin by defining the following super-operator to represent the dynamics generated by the RI map:
\begin{equation}
    \widetilde{\mathcal{H}_n}(\rho) = -i \left[\frac{1}{m} H_0 + H_{E_n} + \frac{1}{\sqrt{\tau}} H_{I_n}, \rho\right].
\end{equation}
This superoperator already accounts for the condition $\lambda^2 \tau \rightarrow 1$ and renormalizes the free Hamiltonian by $m$ to capture the correct dynamics.
\begin{align}
    & \indone{e^{t \liouv} - \Bigr (\bigcirc_{n=1}^m \widetilde{\mathcal{R}_n} (\nicefrac{t}{\nu}) \Bigr )^{\circ \nu}} 
    = \indone{(e^{\frac{t}{\nu} \liouv})^{\circ \nu} - \Bigr (\bigcirc_{n=1}^m \widetilde{\mathcal{R}_n} (\nicefrac{t}{\nu}) \Bigr )^{\circ \nu}}  
     \leq \nu \indone{e^{\frac{t}{\nu} \liouv} - \bigcirc_{n=1}^m \widetilde{\mathcal{R}_n} (\nicefrac{t}{\nu})  } 
     \nonumber\\
    &= \nu \max_{\norm{\rho_S}_1 \leq 1} \Biggr | \Biggr | \Biggr ( \openone -i\frac{t}{\nu}\left[\frac{1}{m}H_0, \cdot\right] - \frac{t}{2\nu}\sum_{k=1}^m\Tr_{\mathscr{H}_E} [H_{I_k}, [H_{I_k}, \cdot \otimes \rho_{E_k}]] + \sum_{j=2}^\infty \frac{\liouv^j t^j}{\nu^j j!} \Biggl ) [\rho_S] \notag \\
    \quad &- \Tr_{\mathscr{H}_E} \prod_{n=1}^m \Biggr ( \openone -i\frac{t}{\nu}\left[ \frac{1}{m} H_0 + H_{E_n} + \sqrt{\frac{\nu}{t}} H_{I_n}, \cdot \otimes \rho_{E_n}\right] - \frac{t}{2\nu}[H_{I_n}, [H_{I_n}, \cdot \otimes \rho_{E_n}]] \notag 
    \nonumber\\
    &- \frac{t^{3/2}}{2\nu^{3/2}} \left( \left[\left(\frac{1}{m} H_0 + H_{E_n}\right), \left[H_{I_n}, \cdot \otimes \rho_{E_n}\right]\right] 
    + \left[H_{I_n}, \left[\left(\frac{1}{m}H_0 + H_{E_n}\right), \cdot \otimes \rho_{E_n}\right]\right]  \right)  \notag \nonumber\\ 
    \quad &- \frac{t^2}{2\nu^2}\left[\left(\frac{1}{m}H_0 + H_{E_n}\right), \left[\left(\frac{1}{m}H_0 + H_{E_n}\right), \cdot \otimes \rho_{E_n}\right]\right] + \sum_{l=3}^\infty \frac{\widetilde{\mathcal{H}_n}^l t^l}{\nu^l l!} 
    \Biggl ) [\rho_S] \Biggl | \Biggl|_1. 
\label{eq:boundRI1}
\end{align}
To ensure convergence of the Taylor series we make necessary smallness assumptions on the Hamiltonian and the interaction time: $\{ \norm{H_0}t/ \nu, \norm{H_E}t/ \nu \} \in o(1)$ and $\norm{H_I} \sqrt{t/ \nu }\in o(1)$. Then, by expanding the product above and tracking only terms $\in O\left(t^2/\nu^2\right)$ we obtain
\begin{align}
    \indone{e^{t \liouv} - \Bigr (\bigcirc_{n=1}^m \widetilde{\mathcal{R}_n} (\nicefrac{t}{\nu}) \Bigr)^{\circ \nu}}   & \in O\Biggr ( \nu \Biggr ( m \max_{n, \norm{\rho_S}_1 \leq 1} \bigr | \bigr | \Tr_{\mathscr{E}_n} \frac{t^2}{2\nu^2}[\frac{1}{m} H_0, [\frac{1}{m} H_0 , \cdot \otimes \rho_{E_n}]] \bigl | \bigl | \notag \\
    \quad &+ \indone{ \sum_{j=2}^\infty \frac{\liouv^j t^j}{\nu^j j!}} + m \max_n \indone{ \sum_{l=3}^\infty \frac{\widetilde{\mathcal{H}_n}^l t^l}{\nu^l l!}} \Biggl ) \Biggr ) \label{eq:fullorder}
\end{align}

This bound was achieved due to the following simplifications:
First, from the first line of Eq. (\ref{eq:boundRI1}) we collected the product of each order of the map with the identity, as these give leading orders in all terms. Then, note that the partial trace over the environment commutes with the free Hamiltonian $H_0$. We also recall that the Lindbladian is  derived with the property $\Tr_{\mathscr{E}_n} [H_I, \rho_S \otimes \rho_{E_n}] = 0$, meaning that the terms $\sqrt{\frac{\nu}{t}} \Tr_{\mathscr{E}_n}[H_{I_n}, \cdot \otimes \rho_{E_n}]$, and all terms that they operate on necessarily vanish. Consequentially, terms above that scale like $O(\frac{t^{1/2}}{\nu^{1/2}})$ and $O(\frac{t^{3/2}}{\nu^{3/2}})$ vanish. Note that the sum in the final term in Equation \ref{eq:fullorder} contains a single term that goes like $O(\frac{t^{3/2}}{\nu^{3/2}})$ due to the inverse square root that multiplies the interaction Hamiltonian. This term involves a triple commutator like so: $\tau^{3/2} [H_{I_n}, [H_{I_n}, [H_{I_n}, \rho ]]]$, however, given the restrictions that impose $\Tr_{\mathscr{E}_n} [H_I, \rho] = 0$, it can be argued that the partial trace over $\mathscr{E}_n$ for any odd order commutator is 0. To make this argument clear, if we are to choose that $\rho_{E_n}$ is a thermal state, as in the case of our prior derivation of Equation \ref{eq:derived_lindblad}, and $H_{I_n}$ also has the form defined in (\ref{eq:derived_lindblad}), then any odd combination of $a$ and $a^\dagger$ makes this necessarily 0, which is a well understood property of creation-annihilation (or raising-lowering) operators. \\
%=================

We now proceed with the bound (\ref{eq:fullorder}), and will revisit $\widetilde{\mathcal{H}}_n^3$ at the end of the proof. 
We proceed by using tail bounds for the two exponential series,
\begin{align}
    \indone{e^{t \liouv} - \Bigr (\bigcirc_{n=1}^m \widetilde{\mathcal{R}_n} (\nicefrac{t}{\nu}) \Bigr )^{\circ \nu}}   & \in O\Biggr ( \nu \Biggr (  \frac{t^2}{m \nu^2} \norm{H_0}^2 +  \frac{\indone{\liouv}^2 t^2}{\nu^2}e^{\frac{t \indone{\liouv}}{\nu}} \notag \\
    &\quad + m  \max_n \Bigr (\frac{t^3\indone{\widetilde{\mathcal{H}_n}^3}}{\nu^3} + \frac{t^4 \indone{\widetilde{\mathcal{H}_n}}^4 }{\nu^4} e^{\frac{t \indone{\widetilde{\mathcal{H}_n}}}{\nu}} \Bigr ) \Biggr ) \Biggr ).
\end{align}
By making the requirements that $\{\frac{t\indone{\liouv}}{\nu} , ~ \frac{t\indone{\widetilde{\mathcal{H}_n}}}{\nu}\} \in [0, \ln{2}] ~ \forall ~ n$, we linearize the bound into the following
\begin{align}
    \indone{e^{t \liouv} - \Bigr (\bigcirc_{n=1}^m \widetilde{\mathcal{R}_n} (\nicefrac{t}{\nu}) \Bigr )^{\circ \nu}}   & \in O \Biggr ( \frac{t^2}{m\nu} \norm{H_0}^2 +  \frac{\indone{\liouv}^2 t^2}{\nu} + m \max_n \Bigr (\frac{t^3\indone{\widetilde{\mathcal{H}_n}^3}}{\nu^2} + \frac{t^4 \indone{\widetilde{\mathcal{H}_n}}^4}{\nu^3} \Bigr )  \Biggr ) .
\end{align}
Now, as a result of prior arguments about odd order commutators with $H_{I_n}$, we have the following 
\begin{align}
    \frac{t^3}{\nu^2}\indone{\widetilde{\mathcal{H}_n}^3} &\in O\Bigr ( \frac{t^3}{m^3 \nu^2} \norm{H_0}^3 + \frac{t^2}{\nu} \norm{H_{E_n}} \norm{H_{I_n}}^2 + \frac{t^2}{m \nu} \norm{H_0} \norm{H_{I_n}}^2 \Bigr ) 
    \nonumber\\
    & \in O \Bigr (\frac{t^3}{m^3 \nu^2} \norm{H_0}^3 + \gamma_n^3 \frac{t^2}{\nu}\Bigr ),
\end{align}
where $\gamma_n \equiv \max \{\norm{H_0}, \norm{H_{E_n}}, \norm{H_{I_n}} \}$. 
Now the first term in the above equation is clearly upper bounded by $\frac{t^2}{m\nu} \norm{H_0}^2$, so we now have 
\begin{align}
    \indone{e^{t \liouv} - \Bigr (\bigcirc_{n=1}^m \widetilde{\mathcal{R}_n} (\nicefrac{t}{\nu}) \Bigr )^{\circ \nu}}   & \in O \Biggr ( \frac{t^2}{m\nu} \norm{H_0}^2 +  \frac{\indone{\liouv}^2 t^2}{\nu} + m \max_n \Bigr (\gamma_n^3 \frac{t^2}{\nu} + \frac{t^4 \indone{\widetilde{\mathcal{H}_n}}^4}{\nu^3} \Bigr )  \Biggr ) .
\end{align}
Now we deal with the final term
\begin{align}
    \frac{t^4 \indone{\widetilde{\mathcal{H}_n}}^4}{\nu^3} & \leq \nu \Bigr (\frac{t}{\nu} \norm{H_0} + \frac{t}{\nu} \norm{H_{E_n}} + \sqrt{\frac{t}{\nu}}\norm{H_{I_n}} \Bigr )^4 \nonumber\\
    & \in O \Biggr ( \nu \Bigr (\sqrt{\frac{t}{\nu}} \norm{H_0} + \sqrt{\frac{t}{\nu}} \norm{H_{E_n}} + \sqrt{\frac{t}{\nu}}\norm{H_{I_n}} \Bigr )^4 \Biggr ) \nonumber\\
    &\in O \Bigr ( \frac{t^2}{\nu}  \gamma_n^4  \Bigr ).
\end{align}
To get the second line we recalled that $\tau = t/\nu$ and that we are interested in the asymptotic limit where $\tau \rightarrow 0$. As such, $\tau<\sqrt \tau$. 
We are now left with 
\begin{equation}
        \indone{e^{t \liouv} - \Bigr (\bigcirc_{n=1}^m \widetilde{\mathcal{R}_n} (\nicefrac{t}{\nu}) \Bigr )^{\circ \nu}}  \in O \Biggr (\frac{t^2}{m\nu} \norm{H_0}^2 + \frac{\indone{\liouv}^2 t^2}{\nu} + m \max_n \Bigr (\gamma_n^4 \frac{t^2}{\nu} \Bigr )  \Biggr ).
        \label{eq:boundRI2}
\end{equation}
Noting that the first term in (\ref{eq:boundRI2}) is asymptotically upper bounded by $m \max_n \gamma_n^4 \frac{t^2}{\nu}$ completes the proof.
    \end{proof}
\end{theorem}

Some useful corollaries immediately follow from the Theorem above:

\begin{corollary}[Number of Interactions] \label{cor:kappa} The minimum necessary number of iterations $\nu$ of an RI map composed of $m$ interactions, written as in Equation (\ref{eq:full_RI}), required to simulate the time evolution of a Lindbladian within error $\epsilon$ has the following upper-bound:
\begin{equation}
    \nu \in O \Biggr ( \frac{t^2}{\epsilon} \Bigr ( \indone{\liouv}^2+ m \max_n \gamma_n^4 \Bigr ) \Biggr ).
\end{equation}
Here $\gamma_n$ is the strength of the interaction used in the $n^{\rm th}$ interaction step.
This then implies that the total number of RI maps $\kappa = m \nu$ that needs to be implemented to realize dynamics $\epsilon$ close to that generated by $e^{t \liouv}$ has the following upper bound:
\begin{equation}
    \kappa \in O \Biggr ( \frac{t^2}{\epsilon} \Bigr ( m\indone{\liouv}^2+ m^2 \max_n \gamma_n^4 \Bigr ) \Biggr ).
\end{equation}
As a direct consequence, this also upper bounds the number of ancilla qubits used in the simulation strategy outlined later in Figure \ref{fig:endtrace}.
\end{corollary}

Furthermore, one of the insights that can be gained through the analysis contained in Theorem \ref{thm:RI_error} is that in terms of the time scaling of the error, the higher orders do not necessarily contribute to converging towards Lindblad dynamics. By this, we mean that by implementing, say terms that come along with $t^3$ or an expanded RI map, we are not improving the accuracy of a Lindblad simulation. Part of the reason is that analyzing the higher order terms is extremely difficult, and potentially intractable using methods from this analysis. In this way, the methods are both poorly sorted to track the higher order contributions, as well as there is no obvious structure or symmetries to indicate higher order convergence. This leads us to conclude that the RI maps defined in this work likely do not lead to higher order integrators, which given the prior discussion in the Introduction, is not of immense concern. In Appendix \ref{sec:appendixC} we perform a short numerical example that supports the conclusions of this Theorem. However, since a large part of the role of $\nu$ is simply to control the higher order terms, it is possible that a cost effective simulation may simply involve not simulating them. We obtain the following Corollary:

\begin{corollary}[Second order Correspondence Error] If an RI map defined as in Equation \ref{eq:full_RI} is implemented up to second order in the interaction time $\tau$, then the induced Schatten $1\rightarrow 1$ norm of the difference between the RI map and the dynamics generated by the Lindblad equation $e^{t\liouv}$ has the following upper bound:
    \begin{equation}
        O \Biggr (\frac{t^2}{m\nu} \norm{H_0}^2 + \frac{\indone{\liouv}^2 t^2}{\nu} \Biggr ),
    \end{equation}
    which can be seen through noting that the $\gamma_n$ terms are derived only from higher order expansions.
\end{corollary}
Given the superior error scaling over Theorem \ref{thm:RI_error}, it may be possible to compose more efficient quantum simulations. However, this truncation leads to a non-unitary map. Implementing this map could therefore be done in a manner that combines strategies proposed in Ref. \cite{cleve2016efficient}. Here, it is argued that a Hamiltonian embedding of the Lindbladian dynamics with repeated trace-out is inefficient compared to the method of using Kraus maps master equations and then implementing them in a purified space using a variant of LCU for isometries. If we were to instead implement repeated interactions to second order, this could be seen as an intermediate approach between these schemes, of which can be made much more compact in the number of qubits in comparison to the Hamiltonian approach in \cite{cleve2016efficient}. However, this then leads to an algorithm where $\delta \neq 1$, and would require a total of $\kappa$ rounds of oblivious amplitude amplification. Therefore, it is unclear how this strategy would compare, a topic of a separate investigation. In this work, we continue to focus on unitary implementations of the RI map using iterative Hamiltonian simulations. \\

To summarize, this section has achieved the goal of bounding the total error between the dynamics generated by a Lindblad equation, and the dynamics realized by $\nu$ applications of an RI map composed of $m$ interactions with an environment. As a corollary, we have derived a bound on the minimum number of system bath interactions to use an RI map to simulate $e^{t\liouv}$ up to precision $\epsilon$. We can now proceed with strategies from quantum computing to analyze the complexity of implementing said repeated interactions, in order to derive the overall complexity of a quantum algorithm based on repeated interactions to simulate Lindbladian Markovian dynamics. 

%====================
\section{RI Implementation with Hamiltonian Simulation}\label{sec:RIsimulation}

In this Section, we write down a unitary form of the map from the previous Section \ref{sec:errorbound}, and show that using the principle of deferred measurement, one can make the entire process unitary prior to tracing out the entire environment. We therefore apply two approaches. In the first one, we iteratively use Hamiltonian simulation strategies to perform the trace, which allows one to bound the number of necessary ancilla qubits by a constant. The other approach preserves unitarity for the entire evolution time, but then requires a number of ancilla that scales polynomially with the simulation time $t$. 
Here, we use Qubitization and Trotter-Suzuki formulas to implement the RI maps. We utilize these algorithms given that they can be performed without amplitude amplification, as both of them have a probability of success $\delta=1$. In addition, Qubitization is utilized to achieve optimal asymptotics, whereas Trotter-Suzuki formulas are chosen for their implementation simplicity, as well as to take advantage of the commutator structure of the problem. The bounds that follow will necessarily depend on $\nu$, however, physically speaking there may exist cases where the RI maps actually provide a more accurate description of the Markovian dynamics than the Lindblad equation, as is likely the case of the maser discussed in Section \ref{sec:intro} and expanded on in Ref. \cite{bruneau2014repeated}. What we mean by this is that the so called \textit{model error} of the Lindblad equation may be greater than that of the RI map, and therefore engineering an RI map to explicitly converge to Lindblad dynamics may be unnecessary. In these cases, the complexity of $\nu$ given in Corollary \ref{cor:kappa} need not apply, and may instead be interpreted as constant. For these reasons, we state our bounds in terms of $\nu$ rather than hashing out the full time complexity, which can simply be obtained by combining the theorems that follow with Corollary \ref{cor:kappa}.

\subsection{Unitary RI Map and Environment Preparation} 
The map $\widetilde{\mathcal{R}}$ derived in the previous section (Eq. \ref{eq:full_RI})
required a Taylor expansion in the interaction time, and then a replacement of the coupling parameter $\lambda \rightarrow 1/\sqrt{\tau}$. This procedure leaves us with a map we can implement to a desired order. However, the map does not have a clear functional form, which can make it difficult to implement with algorithms that are not LCU. The sum form of the map also does not give insight on which parts of the evolution are unitary. Here, we show an RI map that can be written with only unitary operators that produces the same output. \\

\begin{figure}[htbp!]
  \centering
\begin{displaymath}
    \Qcircuit @C=1em @R=1em {
      \lstick{\rho_S} & \qw/^{\log d} & \qw & \multigate{1}{U_1} & \qw & \qw & \qw & \qw & \gate{U_2}  & \qw & \qw & \cdots & ~ & \qw &  \gate{U_\kappa} & \qwa & ~ & ~ &  {\approx \exp(t \liouv(\rho_S))} \\
      \lstick{\rho_{E_1}} & \qw & \qw & \ghost{U_1} & \qw & \qw & \qw & \qw & \qw \qwx & \qw & \qw & \cdots &  & \qw & \qw \qwx & \meter \\
      \lstick{\rho_{E_2}} & \qw & \qw & \qw & \qw & \qw & \qw &\qw & \gate{U_2} \qwx & \qw & \qw & \cdots &  & \qw & \qw \qwx & \meter \\
      \vdots & ~&~ &~ &~ & \vdots  & ~ &~ &~ &~ &  \vdots  & ~ &  ~ &~  &~ \qwx & \vdots \\
      \lstick{\rho_{E_\kappa}} & \qw & \qw & \qw &  \qw & \qw & \qw & \qw &  \qw & \qw & \qw &\cdots & & \qw & \gate{U_\kappa} \qwx & \meter
    }
\end{displaymath}
  \caption{This illustration presents the repeated interactions implementation utilizing the principle of deferred measurement. In this way, the entire process is unitary, availing the use of other quantum algorithms prior to the final measurement, such as amplitude application. The cost of this feature, however, is $\kappa$ qubits, which is upper bounded in Corollary~\ref{cor:kappa}.} \label{fig:endtrace}
\end{figure}

\begin{figure}[htbp!]
  \centering
  \begin{displaymath}
    \Qcircuit @C=1em @R=1em {
      \lstick{\rho_S} & \qw/^{\log d} & \qw & \multigate{1}{U_1} & \qw & \qw & \qw & \qw & \multigate{1}{U_2} & \qw & \qw & \cdots & ~ & \qw & \qw & \multigate{1}{U_\kappa} & \qwa & ~ & ~ &  {\approx \exp(t \liouv(\rho_S))}\\
      \lstick{\rho_{E_1}} & \qw & \qw & \ghost{U_1} &  \meter & ~ & ~ &\lstick{\rho_{E_2}} & \ghost{U_2} & \meter &~& \cdots & ~ & ~ &\lstick{\rho_{E_\kappa}} & \ghost{U_\kappa} & \meter
    }
  \end{displaymath}
  \caption{Illustrated above is the iterative way of applying the repeated interactions map. Given that the complexity of implementing $U_n$ is far greater than preparing the environmental states $\rho_{E_n}$, we can utilize the same register to iteratively prepare our environment and then apply the interaction unitary with our algorithm of choice, to achieve a simulation with a constant number of qubits.} \label{fig:intermittent_trace}
\end{figure}

Suppose we have the unitary $U_n(\tau) = \text{exp}(-i(H_0 + H_{E_n})\tau -iH_{I_n}\sqrt{\tau})$. Lets consider the channel that implements this unitary and an ancillary system to yield the evolution of the reduced dynamics: $\hat{\mathcal{R}}_n[\cdot] = \Tr_{\mathscr{E}_n} U_n (\cdot \otimes \rho_{E_n}) U_n^\dagger$. As before, we also define a super-operator, which in this case is time dependent: $\hat{\mathcal{H}}_n(\rho) = -i[\tau(H_0 + H_{E_n}) + \sqrt{\tau} H_{I_n}, \rho]$. As before, using the adjoint endomorphism of the Lie group, we can write the evolution super-operator $e^{\hat{\mathcal{H}}_n}$. For a RI map with $m$ interactions/ancilla this yields the following:
\begin{equation} \label{eq:unitary_map}
    \bigcirc_{n=1}^m \hat{\mathcal{R}}_n(\rho_S \otimes \rho_{E_n}) = \Tr_{\mathscr{H}_E} \prod_{n=1}^m e^{\hat{\mathcal{H}}_n} (\rho_S \otimes \rho_{E_n})
\end{equation}
Through writing this as a Taylor series of commutators, one can see that this leads to the same error scaling as above (for all orders). This then introduces a subtlety in the algorithmic analysis going forward that is not present in standard Hamiltonian evolution. We have introduced different evolution time scaling for the free and interaction Hamiltonians, ie. without assumptions of the functional scaling of $H_I/ H_0$ or carrying out a more careful analysis with respect to $\tau$, one cannot absorb the spectral norm of $H$ into $t$ and simply apply state of the art bounds. The analyses performed in the following sections reflect this, and carefully track powers of $t$ to modify error bounds from the literature. \\

Through Equation \ref{eq:unitary_map}, we observe that we can perform the RI implementation in one of two ways. We can either perform intermittent trace-out measurements, or performed the trace over the entire environment at the end. The latter is more useful when non-unitary algorithms are used to implement each of the $U_n$s such that amplitude amplification can be used at the end. This method is illustrated in Figure \ref{fig:endtrace}. Given that we have $m$-ancilla qubits to generate the interactions in this case, the number of qubits is naturally upper bounded by Corollary \ref{cor:kappa}. The other approach is a \textit{prepare trace-out refresh} strategy, in which the environment is measured (with the information discarded), and the thermal state is once again prepared or reset. Given that part of the definition of the RI map consists of an environment or ancilla qubit that is tensored on to the system with each application, we must also consider the cost of preparing such a state. Since we are using single qubit thermal states, these can be prepared very efficiently, with complexity far favourable to that of implementing $U_n$. In this way, they can be prepared on the fly, without the need to pre-compute and store ancilla. To show this, we can use Givens rotations strategies to prepare a state vector that is a purification of the thermal state density matrix $\rho_{E_n}$ \cite{nielsen2010quantum}. For an $\eta$-qubit state:
\begin{equation}
    \ket{\psi} = \sum_{j=0}^{2^\eta-1} \frac{\sqrt{\alpha_j} \ket{j} \ket{j}}{\sqrt{\sum_{k=0}^{2^\eta-1} \alpha_k}},
\end{equation}
the preparation has complexity $O(\eta 2^\eta \log (1/ \epsilon))$ in general \cite{kliuchnikov2013synthesis, nielsen2010quantum}. However, given we are initializing a single qubit state, this becomes $O\log (1/\epsilon)$. With this state we simply trace over the second register:
\begin{align}
    \rho_{E_n} &= \Tr_2 (\ketbra{\psi}) \\
    &= \frac{\alpha_0 \ketbra{0}+ \alpha_1\ketbra{1}}{\alpha_0 + \alpha_1},
\end{align}
where the $\alpha_j$s are chosen based on the Givens rotations. In this way, we can efficiently prepare our environment for repeated interactions. For our purposes, when we have a thermal state $\rho_{E_n} = \exp(-\beta a^\dagger a)/\Tr(...)$, and we write $a = X+ iY$, our state has the following coefficients:
\begin{equation}
    \rho_{E_n} = \frac{\ketbra{0} + e^{-\beta}\ketbra{1}}{e^{-\beta}+1}.
\end{equation}

\subsection{Iterative Qubitization of the RI Map}
Since we have been able to write down the RI map as a unitary process, or a process that is at least iteratively unitary with intermediate measurement (depending on where the traces are taken), we can apply state of the art Hamiltonian simulation algorithms to this problem. For qubitization, we wish to write the time evolution as a sum of sine and cosine matrix functions. We can do this here with a redefinition of the interaction Hamiltonian $H_{I_n}^\tau \equiv H_{I_n} / \sqrt{\tau}$. This leaves us with the following:

\begin{equation}
    U_n = \text{exp}(-i\tau(H_0 + H_{E_n} + H_{I_n}^\tau))
\end{equation}

Now, in order to implement this unitary, we construct a block encoding using LCU, which requires that we provide a unitary decomposition of each of our operators here, each of which decomposition has at most polynomially many terms in the number of qubits. We assume the system Hamiltonian $H_0$ has $l_0$ terms and can be written like so (recalling that it has been re-scaled):
\begin{equation}
    H_0 = \sum_{j=1}^{l_0} \frac{h_j}{m} H_j ,
\end{equation}
 where all phase factors are absorbed into $H_j$. We also normalize each of these terms such that $h_j \equiv \norm{H_j}$ and after normalization each $\norm{H_j} = 1$. This can always be done WLOG. Recall, the interaction Hamiltonian was given the earlier definition: $H_I = V_n \otimes a_n^\dagger + V_n^\dagger \otimes a_n$. Here we use the re-scaled $H_{I_n}^\tau$, and also write the general system operator as a sum of unitaries, and the creation-annihilation operators in terms of Pauli operators like so:

 \begin{align}
     H_{I_n}^\tau &= \frac{1}{\sqrt{\tau}} \Biggr ( \sum_{p=1}^{l_{I_n}} v_{p_n} V_{p_n} \otimes \frac{1}{2} (X-iY) + \sum_{q=1}^{l_{I_n}} v_{q_n} V_{q_n}^\dagger \otimes \frac{1}{2} (X+iY) \Biggr ) \\
     & = \frac{1}{\sqrt{\tau}} \Biggr ( \sum_{p=1}^{l_{I_n}} \frac{v_{p_n}}{2} \bigr ( V_{p_n} \otimes X + \hat{V}_{p_n} \otimes Y \bigr ) + \sum_{q=1}^{l_{I_n}} \frac{v_{p_n}}{2} \bigr ( V_{q_n}^\dagger \otimes X + \hat{V}_{q_n}^\dagger \otimes Y \bigr ) \Biggr ), 
 \end{align} 
where we have defined $\hat{V} = -i V$. Finally, we write the free bath Hamiltonians, simply as
\begin{equation}
    H_{E_n} = \omega_n Z,   
\end{equation}
which is the same as the counting operator $a^\dagger a$ up to an energy shift. With these decompositions defined, we can discuss the construction of the block encoding. This can be done more neatly by writing the total Hamiltonian with a re-scaled interaction like so: 
\begin{equation}
    A^\tau_n = H_0 + H_{E_n} + H_{I_n}^\tau = \sum_{s=1}^{l_0+4l_{I_n}+1} \alpha_{s_n} A_{s_n} .
\end{equation}
 Here, the indices work in the order that the respective Hamiltonian matrices were defined above, i.e. $\{ A_{s_n} = V_{s_n} \otimes X ~ : ~ s \in [l_0 +1, l_0 + l_{I_n}] \}$. With this definition, the block encoding machinery can be explained more clearly. Our construction is similar to that of Ref. \cite{Berry_2015} We define the \textit{prepare} operator $P$ as a unitary to load the necessary coefficient. This works in the following way
 \begin{equation} \label{eq:prep}
     P_n \ket{0}^{\otimes c} = \frac{1}{\sqrt{\alpha_n}}\sum_{\{s_{n}\}} \sqrt{\alpha_{s_n}} \ket{s_n},
 \end{equation}
where $P_n$ is a unitary of dimension $c = l_0+4l_{I_n}+1$, and $\alpha_n = \sum_{\{s_{n}\}} \alpha_{s_n}$ for normalization. We will refer to this sum of coefficients as the alpha norm, which we be an important factor in the final bound we present in the Section. Next, we build a \textit{select} operator $S_n$ which applies a certain Hamiltonian term conditioned on the superposition state produced by $P_n$. The operator $S_n$ acts in the following way:
\begin{equation} \label{eq:select}
    S_n \ket{s_n} \ket{\psi} = \ket{s_n} A_{s_n} \ket{\psi} .
\end{equation}
Here $\ket{s_n}$ represents our control state register, while $\ket{\psi}$ represents our input quantum state. In general, our input state is a mixed quantum state $\rho$, but given the representation of the map in Equation \ref{eq:unitary_map}, this formulation is equivalent, and writing all of these operations as channels makes the derivation unnecessarily complex. To remind the reader, the $n$ subscript is allowing for the freedom to define different interactions between the environment and system as well as the bath frequency $\omega_n$. For this reason, a different choice of interaction may yield a different unitary decomposition that requires a prepare operator $P$ that has a different Hilbert space dimension. For this reason, we continue to subscript our \textit{prepare} and \textit{select} unitaries.\\

With the above machinery defined, we now show how to utilize it to prepare a block encoding:
\begin{equation}
    (P_n^\dagger \otimes \openone_d) S_n (P_n \otimes \openone_d) \ket{0}^{\otimes c} \ket{\psi} = \ket{0}^{\otimes c} \frac{A^\tau_n}{\alpha_n}\ket{\psi} + \ket{\phi^\perp},
\end{equation}
where $\ket{\phi^\perp}$ is a properly normalized state that is not of interest, and $d=\dim \mathscr{H}_0$. With the above operation, we have successfully produced the following block encoding of the matrix $A_n^\tau$:
\begin{equation}
  (P_n^\dagger \otimes \openone_d) S_n (P_n \otimes \openone_d) = \scalebox{1.5}{
    $\begin{bmatrix}
      \frac{A_n^\tau}{\alpha_n} & \cdot \\
      \cdot & \cdot \\
    \end{bmatrix}$
  },
\end{equation}
which can be accessed in the following way:
\begin{equation}
    (\bra{0}^{\otimes c} \otimes \openone_d)  (P_n^\dagger \otimes \openone_d) S_n (P_n \otimes \openone_d) (\ket{0}^{\otimes c} \otimes \openone_d ) = \frac{A_n^\tau}{\alpha_n}.
\end{equation}
The above example illustrates that the block encoding can be accessed upon measurement of the $\ket{0}^{\otimes c}$. In implementations of LCU strategies, such as in Ref. \cite{Berry_2015}, it is required that one uses the amplitude amplification algorithm to boost the probability of success $\delta$ of this measurement. However, in our strategy we are interested in methods with $\delta = 1$. Therefore, we use Qubitization and QSVT to simulate our RI maps which implement via queries to $P$ and $S$ defined above. The number of queries $q_{PS}$ to the \textit{prepare} and \textit{select} oracles required to simulate $\liouv$ with $\mathcal{R}$ up to precision $\epsilon$ is bounded in the following Theorem.
\begin{theorem}[Iterative Qubitization Complexity] \label{thm:qubitization}
Given a RI map from Definition \ref{def:RI}, an initial state $\rho_S$ and total time $t$, the time evolution $\exp(\liouv t)$ of the corresponding Lindbladian from Equation \ref{eq:rescaled_Lindblad} can be implemented with the following number of queries $q$, to both oracles $P$ from Equation \ref{eq:prep} and $S$ from Equation \ref{eq:select}:
\begin{equation}
     q_{PS} \in O\Biggr ( \max_n \biggr((\alpha_0 + m \omega_n)t +  \alpha_{I_n} m \sqrt{t \nu} + \frac{m \nu \log \frac{1}{\epsilon}}{\log \log \frac{1}{\epsilon}} \biggr ) \Biggr ),
\end{equation}
with $\nu$ applications of an RI map composed as in Equation \ref{eq:full_RI}, $m$ environment/bath ancilla, $\omega$ the bath frequency, and $\alpha$ the coefficient 1-norm for the system and interaction Hamiltonians.
\begin{proof}
        
The art of qubitization requires us to define a reflection operator about the 2-dimensional subspace in which the \textit{prepare state} is embedded. This can be achieved as follows:
\begin{equation}
    R_n = (\openone - 2  (P_n \otimes \openone_d) \ket{0}\bra{0} (P_n^\dagger \otimes \openone_d)).
\end{equation}
With this reflection, we construct the walk operator below, which is used in the implementation of the dynamics of the $n$th subsystem
\begin{equation} \label{eq:walk_op}
    W_n = R_n  S_n .
\end{equation}
Qubitization bounds are formulated in terms of the number of queries to the walk operator, however, for better interpretability, one can see that this requires 1 query to \textit{select} $S_n$ and 2 queries to \textit{prepare} $P_n$, which is equivalent in terms of asymptotics. The Quantum Singular Value Transformation then provides the machinery for using this walk operator to implement the operator exponential using a series of $X$ and $Z$ rotations in the 2-dimensional subspace. To see this discussion, consult Refs \cite{gilyen2019quantum, martyn2021grand}. The aforementioned analysis leads to the following complexity upper-bound:
\begin{equation} \label{eq:qsvt_bound}
    q_{PS} \in O \Biggl ( \alpha t + \frac{\log \frac{1}{\epsilon}}{\log \log \frac{1}{\epsilon}} \Biggr ),
\end{equation}
where $t$ is the total simulation time and $\alpha$ is the sum of the coefficients from the unitary decomposition previously discussed. In the case of the RI map, we can use this upper bound almost directly, with a few minor adjustments. First, we are simulating each individual application of the map with this algorithm a total of $\kappa$ times, meaning we make the replacement $t \rightarrow \tau$, and we pick up a multiplicative factor of $\kappa$. Recall from Corollary \ref{cor:kappa} that $\kappa = m\nu$ is the total number of repeated interactions in the composite map defined in Equation \ref{eq:full_RI}. Also, given our definition of $A_N^\tau$, the $\alpha$-norm argument goes through slightly differently. For RI maps, the component of the $\alpha$-norm that comes from the interaction Hamiltonian decomposition is necessarily a function of time, given that the Hamiltonian has been re-scaled to $H_{I_n}^\tau$. It is cleaner to instead leave it as a multiplicative factor, and we do the same for the re-scaled free Hamiltonian $H_0$: 

\begin{align}
    \alpha_n &= \frac{1}{m} \alpha_0 + \sqrt{\frac{1}{\tau}} \alpha_{I_n} + \alpha_{E_n} \\
    & = \sqrt{\frac{\nu}{t}} \alpha_{I_n} + (\frac{1}{m}\alpha_0 + \omega_n).
\end{align}
 Now, applying Equation \ref{eq:qsvt_bound} and multiplying by a factor of $m$, and inserting our defined $\alpha$-norms, we obtain the following bound:

 \begin{equation}
     q_{PS} \in O\Biggr ( \max_n \biggr((\alpha_0 + m \omega_n)t +  \alpha_{I_n} m \sqrt{t \nu} + \frac{m \nu \log \frac{1}{\epsilon}}{\log \log \frac{1}{\epsilon}} \biggr ) \Biggr ).
 \end{equation}
 This completes the proof.
 \end{proof}
 \end{theorem}

Recall that Corollary \ref{cor:kappa} places an upper bound on $\nu$ to give the overall time complexity.

\subsection{Trotter-Suzuki Formulas}
In this Section we implement the RI map using the well known Trotter-Suzuki product formulas \cite{suzuki1990fractal, childs2021theory}, and discuss some technicalities that arise as a result of redefining the coupling $\lambda \rightarrow 1/\sqrt{\tau}$. Our motivation for using these formulas are again that we can implement the Lindbladian dynamics with $\delta = 1$, for means of comparison with Ref. \cite{cattaneo2021collision}, which first implemented a variant of these maps with Trotterization. For an introduction to these product formulas, see Ref. \cite{Hatano_2005}, or the preliminaries contained in Ref. \cite{pocrnic2024composite}. Given that we have some exponentiated sum of operators $e^{x\sum_i A_i}$ and we define the 1st and 2nd order product formulas as follows:

\begin{align} 
    \trotterchan{1}{Ax} &= \prod_i^L e^{A_i x}  \label{eq:trot1} \\
    \trotterchan{2}{Ax} &= \prod_i^L e^{\frac{A_i x}{2}} \prod^i_L e^{\frac{A_i x}{2}} \label{eq:trot2} 
\end{align}

we can define a $2k$th-order product formula in the following way:

\begin{equation}
    \trotterchan{2k}{Ax}:= (\trotterchan{2k-2}{Axs_{2k}})^2 \trotterchan{2k-2}{(1-4s_{2k})Ax} (\trotterchan{2k-2}{Axs_{2k}})^2 ,
\end{equation}

where $s_{2k} = \frac{1}{4 - 4^{\frac{2k-1}{2}}}$. Then to generalize this to density matrices we can define a Trotter channel in the following way:

\begin{definition}[Trotter-Suzuki Channel] \label{def:Trotter_channel} Given a Hamiltonian $H$, a density matrix $\rho$, times $t$ and $t_0$ ($t> t_0 \geq 0$), and order 2k, then a Trotter-Suzuki channel $\tschan{2k}{\rho(t_0), \Delta t}$ performs the operation $\tschan{2k}{\rho(t_0), \Delta t} \rightarrow \rho(t)$ and can be defined as:
\begin{equation}
    \tschan{2k}{\rho(t_0), t} := \trotterchan{2k}{-iH \Delta t} \rho \trotterchan{2k}{-iH \Delta t}^\dagger
\end{equation}
Where $\trotterchan{2k}{-iHt}$ represent the product formulae from Equation (\ref{eq:trot2}) and $\Delta t = t-t_0$.
\end{definition}
To control the error in $t$, it is common to implement the channel with for time $t/r$ and iterate $r$ times. With these channels defined, we can now bound the error of a RI Trotter implementation and make conclusions about the cost of such a simulation. However, the input model in this case is different, considering that we do not need to construct any block encodings, but rather we implement the unitaries directly. To make this precise, given that we are provided with a Hamiltonian $H_n$ at each interaction that is written as a sum of operators $H_n = H_0 + {H_{I_n}+H_{E_n}} = \sum_{j=1}^l H_j$ , with $l \in \text{poly}(\eta)$ where $\eta$ is the number of qubits. Now we consider the cost of the algorithm as simply the number of exponential gates needed to implement a product formula. This can likewise be thought of as queries $q_{TS}(2k)$ to simple quantum circuits that provide the operator of the Hamiltonian summands $e^{-iH_j t/r}$ with the desired time slice. In most common applications of said decompositions, where we write our Hamiltonian as a sum of Pauli operators or fermionic creation and annihilation operators, simple circuits for such a task are well understood \cite{whitfield2011simulation}. Note $2k$th order product formulas contain $\Upsilon$ stages, where
\begin{equation}
    \Upsilon \equiv 2 \cdot 5^{k-1},
\end{equation}
and a stage can be though of as a string that contains each of the $l$ Hamiltonian summands $H_j$. Therefore, if we have time-slice $t/r$ and we implement the entire product formula $r$ times, it is equivalent to carrying out $q_{TS}(2k) = l \Upsilon r$ gates, where for a first order formula we simply have 1 stage. With this in mind, we derive the following bounds on $q_{TS}(2k)$ for a RI map that implements Lindbladian dynamics:

\begin{theorem}[Higher Order Trotter-Suzuki Cost] \label{thm:Trotter2k}
Given a RI map from Definition \ref{def:RI}, an initial state $\rho_S$ and total time $t$, the time evolution $\exp(\liouv t)$ of the corresponding Lindbladian from Equation \ref{eq:rescaled_Lindblad} can be implemented with a $2k$th order Trotter-Suzuki channel $\tschan{2k}{t}$ from Definition \ref{def:Trotter_channel}, using the following number of exponential gates $q_{TS}(2k)$:
\begin{equation}
    q_{TS}(2k) \in O \Biggr ( l \Upsilon \biggr(  \frac{ \max_n \Bigr ( (\alpha_0 + m \omega_n)t + m \alpha_{I_n} \sqrt{\nu t} \Bigr )^{1+\frac{1}{2k}} }{m \nu \epsilon^{\frac{1}{2k}}}  \biggr) \Biggr ).
\end{equation}
    \begin{proof}
        The procedure consists of bounding the error we define by the induced one norm, setting this error equal to $\epsilon$ and rearranging for $r$. Equipped with an upper bound on $r$, we can then upper bound the total number of gates using $q = l \Upsilon r$ as discussed above. Throughout, we will index the channels such that the indexing sets have the following structure $\{\{n=1, ..., n =m\} \times \nu \}$. This makes the RI map written below equivalent to that defined in Eq. \ref{eq:full_RI} and leads to more straightforward applications of inequalities:
\begin{align}
    \indone{\bigcirc_{n=1}^{m\nu} \mathcal{T}_n^{2k}(t/\nu)  -  e^{t \liouv}} &\leq 
    \indone{\bigcirc_{n=1}^{m\nu} \mathcal{T}_n^{2k}(t/\nu) - \bigcirc_{n=1}^{m\nu} \widetilde{\mathcal{R}_n}(t /\nu)} + \indone{\bigcirc_{n=1}^{m\nu} \widetilde{\mathcal{R}_n}(t/\nu) - e^{t \liouv}} \\
    &\leq m\nu \max_n \indone{\mathcal{T}_n^{2k}(t / \nu) - \widetilde{\mathcal{R}_n}(t/ \nu)} + \epsilon_{RI} \label{eq:trotterRI_error}
\end{align}

In the second line, we utilize Lemma IV.2 from Ref. \cite{pocrnic2024composite}, a minor channel composition inequality. We also observe that the induced 1-norm on the first line is the RI error that has already been bounded previously. From here, we can apply Theorem 2 from \cite{hagan2022composite} to get the form of the Trotter error. The aforementioned Theorem generalizes bounds from Ref. \cite{childs2021theory} to channels. In the case of arbitrarily high order Trotter-Suzuki formulas, an analysis of the commutator structure is highly non-trivial. In general, given the error scaling of $m$, and the accuracy or \textit{model error} of the Lindblad equation itself, higher order Trotter formulas may not be as useful as they are other applications. For these reasons, we provide a trivial upper bound for the $2k$th order formulas by implementing Lemma 1 from Ref. \cite{childs2021theory}.

\begin{align}
     \indone{\bigcirc_{n=1}^{m\nu} \mathcal{T}_n^{2k}(t/\nu)  -  e^{t \liouv}} &\in 
     O \Biggr ( m\nu \max_n \biggr(  \frac{\alpha_n^{2k+1} (t/\nu)^{2k+1}}{r^{2k}}  \biggr) + \epsilon_{RI} \Biggr ) \\
     & \in O \Biggr ( m \max_n \biggr(  \frac{(\alpha_0/m + \alpha_{I_n} \sqrt{\nu/t} + \omega_n)^{2k+1} t^{2k+1}}{r^{2k} \nu^{2k}}  \biggr)  + \epsilon_{RI} \Biggr )
\end{align}

In both of the cases above, for the purposes of asymptotic analysis, we can make the promise that the total simulation error $\epsilon = \frac{1}{2} \epsilon_{RI}$, to avoid solving a non linear equation. With this, we can simply rearrange the above bounds for $r$:

\begin{align}
     r & \in O \Biggr ( \max_n \biggr(  \frac{\Bigr ( (\alpha_0 + m \omega_n)t + m \alpha_{I_n} \sqrt{\nu t} \Bigr )^{1+\frac{1}{2k}} }{m \nu \epsilon^{\frac{1}{2k}}}  \biggr) \Biggr ),
\end{align}
and and multiplying by $l \Upsilon$ completes the proof.
    \end{proof}
\end{theorem}

For simplicity, lets also consider the case of the first-order Trotter-Suzuki channel. Give the more simple commutator sums that appear in the Trotter error for low order formulas, we utilize some of the known commutator structure to give more detailed bounds in the following Theorem.

\begin{theorem}[Trotterization Cost] \label{thm:first_trotter}
Given a RI map from Definition \ref{def:RI}, an initial state $\rho_S$ and total time $t$, the time evolution $\exp(\liouv t)$ of the corresponding Lindbladian from Equation \ref{eq:rescaled_Lindblad} can be implemented with a 1st order Trotter-Suzuki channel $\tschan{2k}{t}$ from Definition \ref{def:Trotter_channel}, using the following number of exponential gates $q_{TS}(1)$:
\begin{equation}
   q_{TS}(1)  \leq l \Biggr ( \max_n \frac{t^{3/2}}{2\epsilon \sqrt{\nu}} \Bigr (\sum_{j=1}^{l_0} \norm{[H_{0_j}, H_{I_n}]} + m \norm{[H_{E_n}, H_{I_n}]} \Bigr) +  \frac{ t^2}{2 \epsilon \nu} \sum_{p=1, q=1}^{l_0, l_0} \frac{1}{m} \norm{[H_{0_p}, H_{0_q}]} \Biggr )
\end{equation}
    \begin{proof}
Starting from Equation \ref{eq:trotterRI_error}, we can once again apply Theorem 2 from \cite{hagan2022composite} to insert the form of the Trotter error. Note that given the low order of the formula, we can more simply avail some of the commutator structure in the RI map. We have in general $[H_0, H_{I_n}^\tau] \neq 0$, $[H_{E_n}, H_{I_n}^\tau] \neq 0$, and $[H_{E_n}, H_0]=0$. 
   \begin{align}
    &\indone{\bigcirc_{n=1}^{m\nu} \mathcal{T}_n^{1}(t/\nu )  -  e^{t \liouv}} \leq m \nu \max_n \indone{\mathcal{T}_n^{1}(t/ \nu) - \widetilde{\mathcal{R}_n}(t)} + \epsilon_{RI} \\
    &\leq m \max_n \frac{t^2}{2 \nu r} \Bigr (\sum_{j=1}^{l_0} \frac{1}{m}\norm{[H_{0_j}, H_{I_n}^\tau]} + \norm{[H_{E_n}, H_{I_n}^\tau]} + \sum_{p=1, q=1}^{l_0, l_0} \frac{1}{m^2}\norm{[H_{0_p}, H_{0_q}]} \Bigr)+ \epsilon_{RI} \\
    & =  \max_n \frac{t^{3/2}}{2r\sqrt{\nu}} \Bigr (\sum_{j=1}^{l_0}  \norm{[H_{0_j}, H_{I_n}]} + m \norm{[H_{E_n}, H_{I_n}]} \Bigr) +  \frac{ t^2}{2r\nu} \sum_{p=1, q=1}^{l_0, l_0} \frac{1}{m} \norm{[H_{0_p}, H_{0_q}]} + \epsilon_{RI} 
\end{align}
    where in the previous line we pulled the factor of $1/\sqrt{\tau}$ out of the interaction Hamiltonian. Once again making the promise that $\epsilon = \frac{1}{2} \epsilon_{RI}$, and rearranging for $r$ we obtain the following:
\begin{equation}
    r \leq  \max_n \frac{t^{3/2}}{2\epsilon \sqrt{\nu}} \Bigr (\sum_{j=1}^{l_0} \norm{[H_{0_j}, H_{I_n}]} + m \norm{[H_{E_n}, H_{I_n}]} \Bigr) +  \frac{ t^2}{2 \epsilon \nu} \sum_{p=1, q=1}^{l_0, l_0} \frac{1}{m} \norm{[H_{0_p}, H_{0_q}]},   
\end{equation}
now we can trivially bound the number of operator exponentials $q_{TS}(1)$ by multiplying by the total number of terms in the Hamiltonian $l$ which completes the proof.
    \end{proof}
\end{theorem}

As in the case of the iterative Qubitization simulation, Corollary \ref{cor:kappa} can be applied to upper bound the number of applications of the composed RI map defined in Equation \ref{eq:full_RI}.

\section{Discussion}
In this work, we successfully bounded the error between a Lindbladian evolution and a Repeated Interaction map in Theorem \ref{thm:RI_error}. In doing so, we showed the the error in this approximate correspondence can be made arbitrarily small by increasing the number of interactions between the quantum system of interest and environmental subsystems. This also implies that dissipative Lindbladian dynamics can be efficiently embedded as a sequence of unitary Hamiltonian simulations with trace out. As a consequence, this analysis also places an upper bound on the number of ancilla qubits needed, given that we perform a deferred trace out to maintain unitarity for the entire simulation time $t$. We then implemented iterated Qubitization and QSVT strategies to yield an asymptotically optimal implementations of RI maps, as well as two Trotter-Suzuki simulations for comparison. Complexity upper bounds of these approaches were given in Theorems \ref{thm:qubitization},  \ref{thm:Trotter2k}, and \ref{thm:first_trotter}. All algorithms derived in this work have complexity at least as efficient as what is presented in Ref. \cite{cattaneo2021collision}, with qubitization and higher order Trotter-Suzuki formulas doing even better. This is not achieved by circumnavigating the $t^2$ dependence of the Lindblad-RI correspondence error $\epsilon_{RI}$, rather by shifting a large proportion of simulation parameters to multiply weaker powers of $t$. In this manner, the utilization of a qubitization approach provides what is to our knowledge the best known scaling for simulating Lindbladian dynamics with RI. However, in terms of time complexity, the nature of the convergence parameter $\nu$ limits this approach from achieving what is asymptotically optimal for outright Hamiltonian simulation. For example, considering the time complexity of $\nu$, our qubitization approach achieves $O(t^{3/2} + t^2 \text{polylog}(1/\epsilon))$, and it is unclear whether this can be further improved. Therefore, the question of whether Hamiltonian simulation lower-bounds can be saturated in Lindbladian simulation remains open. \\

What is somewhat unique to this problem, is that even in generality, some commutator structure is known a priori. This potentially makes for a good candidate for composite simulation strategies such as those discussed in Refs. \cite{hagan2022composite, pocrnic2024composite}. Specifically, in Ref. \cite{hagan2022composite} it was proven that if one can find a partition $H=A+B$ such that the number of non-zero commutators in $A$ scales like the square root of the number of terms in the set $|A|$, and the terms in $B$ are sufficiently small, then it is always possible to gain an asymptotic advantage by simulating $A$ with Trotter and $B$ with QDrift. Whether the RI map guarantees the construction of the aforementioned sets is an open question (see Section 4.3 of \cite{hagan2022composite}).\\

This problem is also interesting due to the two errors that are coming in, one from the approximate correspondence between Lindblad and RI dynamics, and the other from the quantum simulation. Interestingly, the asymptotics suggest that implementing high orders of the RI map are not guaranteed to reduce $\epsilon_{RI}$. This means that iteratively implementing just the first two orders of the map, in a non-unitary fashion, may lead to a more efficient simulation. For instance, constructing a Kraus map that approximates the first two orders of $\widetilde{\mathcal{R}_n}$ and then applying an LCU strategy, similar to the approach of \cite{cleve2016efficient}, may outperform the prior approach when it comes to constant factors. However, our goal in this work was to provide unitary algorithms based on Hamiltonian simulation with $\delta=1$, so we leave this approach as a future avenue to be explored. \\

Another interesting avenue is exploring different RI schemes in the context of simulation. This investigation was particularly done under the condition $\lambda^2 \tau = 1$, with a trace and refresh after every interaction. It is without question that this approach can be extended to non-Markovian dynamics simply by not refreshing the bath, but this then leads to the question of ``to which quantum master equation are we converging too?". In terms of the relationship between coupling and interaction time, other master equations may potentially also be implemented in accordance with other relationships investigated in Ref. \cite{attal2007weak}. \\

From a broader perspective, our work reveals an issue with current approaches to simulating open system dynamics.  Often the master equations that we aim to simulate have fundamental \textit{model errors} that make the dynamics only approximately equivalent to the exact unitary dynamics that describes a system bath coupling.  Understanding these errors in the context of open systems simulation is important since bounding such discrepancies may be vital to fairly compare different open systems simulation schemes.  In particular, a complete closed system simulation for a system may only be an approximation to Lindbladian dynamics but it may be a more accurate description of the quantum dynamics than the Lindblad master equation because it does not make the Markov approximation.  While our work makes an important step in this direction, we feel more work is needed in order to understand how best to use approaches from open quantum systems to simulate quantum dynamics with provable error guarantees, while also retaining the conceptual simplicity and clarity that the open systems approach provides.

\section{Acknowledgements}
We thank Matthew Hagan, Danial Motlagh, and Marlon Brenes for valuable discussions about this work as well as Lin Lin, Xiantao Li, and Shantanav Chakraborty for useful feedback on our work.  MP and NW acknowledge funding from the NSERC Discovery Program.  NW's work was further supported by the US Department of Energy, Office of Science, National Quantum Information
Science Research Centers, Co-Design Center for Quantum Advantage under contract number
DE-SC0012704. DS acknowledges support from an
NSERC Discovery Grant and the Canada Research Chair program. 

\section{Conflicts of Interest}
The authors declare no conflicts of interest.

\section{Data Availability Statement}
No new data was created or analyzed in this study.

\bibliography{bib} 

\appendix
\section{General Error Bounding} \label{sec:appendixA}
This Section is dedicated towards bounding the following induced 1-norm $\norm{e^{\tau \mathcal{L}} - \mathcal{R}_n (\tau)}_{1 \rightarrow 1}$. In order to do this, and for simplicity, we assume that the Lindblad generator is exactly equivalent to the repeated interaction Lindblad form derived in Section \ref{sec:lindblad_derivation}. This will allow for the cancellation of some terms in the series expansion. Rather than yielding the best overall bound in this work, this section is written to stress the importance of the limiting case of the map. Note that we are here working with the general RI map $\mathcal{R}_n (\tau)$ rather than the limiting case $\widetilde{\mathcal{R}_n} (\tau)$ % DS
used in the Lindbladian derivation in the main text. The overall error expression between the Lindblad dynamics and the General RI is give by
\begin{align} \label{error_definitions}
    \norm{e^{\tau \mathcal{L}} - \mathcal{R}_n (\tau)}_{1 \rightarrow 1} \leq \underbrace{\indone{e^{\tau \mathcal{L}} - \widetilde{\mathcal{R}_n} (\tau)}}_{\epsilon_\tau} + \underbrace{{\indone{\widetilde{\mathcal{R}_n} (\tau) - \mathcal{R}_n (\tau)}}}_{\epsilon_\lambda}.
\end{align}
Here $\widetilde{\mathcal{R}_n} (\tau)$ is the repeated interaction map in the $\lambda^2 \tau \rightarrow 1$ case, or, when expanded, the map that leads to Equation (\ref{eq:lambdalim}). We also use the notation $\widetilde{\mathcal{H}_n}= -i[H_0 + H_{E_n} + \frac{1}{\sqrt{\tau}}H_{I_n}, \cdot ]$ to be the Hamiltonian super-operator in this same case. Here, we split the error into two terms, $\epsilon_\tau$ for the error that arises from the fact that $\tau$ does not precisely go to zero in the repeated interactions map, and the other $\epsilon_\lambda$ which is the difference arising from $\lambda$ not going to $\sqrt{\nicefrac{1}{\tau}}$ in general. Note that if we are dealing with an RI map composed of the full $m$ interactions, then $\epsilon_{\tau} = \epsilon_{RI}$, or the error bounded in Theorem \ref{thm:RI_error}. Therefore, we are calculating the error between a Lindblad evolution and a general repeated interactions map with no additional assumptions:
\begin{align}
    \epsilon_\tau &= \indone{e^{\tau \mathcal{L}} - \widetilde{\mathcal{R}_n} (\tau)} \\
    &= \indone{\int_0^\tau \partial_x \parens{e^{x \mathcal{L}} - \widetilde{\mathcal{R}_n} (x)} dx + \parens{e^{0 \mathcal{L}} - \widetilde{\mathcal{R}_n} (0)}} \\
    &= \indone{\int_0^\tau \partial_x \parens{e^{x \mathcal{L}} - \widetilde{\mathcal{R}_n} (x)} dx} \\
    &\leq \int_0^\tau \indone{ \partial_x \parens{e^{x \mathcal{L}} - \widetilde{\mathcal{R}_n} (x)}}dx \\
    & \leq \tau \max_{x \in [0, \tau]} \Biggr | \Biggr |
     \mathcal{L} \sum_{n=1}^\infty \frac{\mathcal{L}^n x^n}{n!} - \Tr_{\mathscr{E}_n} \Biggl (\partial_x \sum_{m=3}^\infty \frac{\widetilde{\mathcal{H}_n}^m x^m}{m!} \Biggr ) \Biggr | \Biggr |_{1\rightarrow 1} \notag \nonumber\\
     &\quad + \max_{\rho : \norm{\rho}_1 = 1} \Biggr | \Biggr | \Tr_{\mathscr{E}_n} \Biggr (\frac{3}{4} \tau^{1/2} \Bigl ([(H_0 + H_{E_n}), [H_{I_n}, \rho_S \otimes \rho_{E_n}]]  [H_{I_n}, [(H_0 + H_{E_n}), \rho_S \otimes \rho_{E_n}]] \Bigr ) + \notag 
     \nonumber\\ 
     & \quad+ \tau [(H_0 + H_{E_n}),[(H_0 + H_{E_n}),\rho_S \otimes \rho_{E_n}]] \Biggr ) \Biggr | \Biggr | 
\end{align}
Looking at the above result, we simplify it by using the fact that $\Tr_E [H_E, \rho_S \otimes \rho_{E_n}] = 0$ and (based on the assumption in the derivation) $\Tr_E [H_I, \rho_S \otimes \rho_{E_n}] = 0$, as well as the fact that $H_0$ commutes with the partial trace, we have that 
\begin{equation}
\Tr_{\mathscr{E}_n} \frac{3}{4} \tau^{1/2} \Bigl ([(H_0 + H_{E_n}), [H_{I_n}, \rho_S \otimes \rho_{E_n}]] + [H_{I_n}, [(H_0 + H_{E_n}), \rho_S \otimes \rho_{E_n}]] \Bigr ) = 0.  
\end{equation}
After applying the triangle inequality and making some smallness assumptions, we can see that, to leading order we will have:
\begin{equation}
    \epsilon_\tau \in O(\tau^2 \indone{[(H_0 + H_{E_n}),[(H_0 + H_{E_n}),\rho_S \otimes \rho_{E_n}]]}).
\end{equation}
This term contributes an error that scales like $\tau^2$, which without additional assumptions seems to be the best one can approximate Lindbladian evolution with an RI map. We will make these arguments more rigorous in the next section.
However, this is not the case with the other remaining error term $\epsilon_\lambda$. Turning our focus to $\epsilon_\lambda$, and simply expanding each map and cancelling like terms we have: 

\begin{align}
    \epsilon_\lambda &= \indone{\widetilde{\mathcal{R}_n} (\tau) - \mathcal{R}_n (\tau)} \\
    &= \max_{\rho : \norm{\rho}_1 = 1} \Biggr | \Biggr | \Tr_{\mathscr{E}_n} \Biggl(\sum_{k=3}^\infty \frac{\mathcal{H}_n^k x^k}{k!} - \sum_{m=3}^\infty \frac{\widetilde{\mathcal{H}_n}^m x^m}{m!}  + \frac{1}{2}\Bigl(  (\tau^2 \lambda - \tau^{3/2})([H_{I_n}, [(H_0 + H_{E_n}), \rho_S \otimes \rho_{E_n}]] \notag \\ 
    &\quad + [(H_0 + H_{E_n}), [H_{I_n}, \rho_S \otimes \rho_{E_n}]]) + (\tau^2 \lambda^2 - \tau)[H_{I_n}, [H_{I_n}, \rho_S \otimes \rho_{E_n}]] \Bigr )  \Biggr ) \Biggr | \Biggr |_1
\end{align}
We can proceed with repeated applications of the triangle inequality and sub-multiplicative property. We will also utilize the von Neumann trace inequality to bound the partial traces. Furthermore, we utilize the property 
\begin{equation}
    \indone{\mathcal{H}} \leq 2\norm{H} \leq 2(\norm{H_0} + \norm{H_{E_n}} + \lambda \norm{H_{I_n}})
\end{equation}
This bound is saturated when the density matrix support is constrained to the 1-D subspace that coincides with the largest eigenvalue of the Hamiltonian. Proceeding with applications of the triangle inequality we have:
\begin{align}
    \epsilon_\lambda &\leq \indone{\Tr_{\mathscr{E}_n} \sum_{k=3}^\infty \frac{\mathcal{H}_n^k x^k}{k!}} + \indone{ \Tr_{\mathscr{E}_n}\sum_{m=3}^\infty \frac{\widetilde{\mathcal{H}_n}^m x^m}{m!}} \notag \\
    &\quad+ \frac{1}{2}\Bigl(  (\tau^2 \lambda + \tau^{3/2}) (\indone{ \Tr_{\mathscr{E}_n} [H_{I_n}, [(H_0 + H_{E_n}), \rho_S \otimes \rho_{E_n}]]} \notag \\
    &\quad+ \indone{ \Tr_{\mathscr{E}_n}[(H_0 + H_{E_n}), [H_{I_n}, \rho_S \otimes \rho_{E_n}]]}) + (\tau^2 \lambda^2 + \tau) \Tr_{\mathscr{E}_n} \indone{ [H_{I_n}, [H_{I_n}, \rho_S \otimes \rho_{E_n}]]} \Bigr ),
\end{align}
Immediately we notice something problematic with this error. The following term, 
\begin{equation}
    (\tau^2 \lambda^2 + \tau) \Tr_{\mathscr{E}_n} \indone{ [H_{I_n}, [H_{I_n}, \rho_S \otimes \rho_{E_n}]]} \neq 0, \: \forall \tau > 0.
\end{equation}
as we know the double commutator with the interaction Hamiltonian to necessarily be non-zero \cite{bruneau2014repeated}. Further bounding this term, we make the assumption on the form of the interaction Hamiltonian $H_{I_n} = S \otimes B_n$, 
\begin{align}
    \Tr_{\mathscr{E}_n} \indone{ [H_{I_n}, [H_{I_n}, \rho_S \otimes \rho_{E_n}]]} & \leq \abs{\Tr_{\mathscr{E}_n} (B_n^2 \rho_{E_n})} \indone{S^2\rho_S + \rho_S S^2 - 2 S \rho_S S} \\
    & \leq \sum_j e_j(B^2) e_j(\rho_{E_n}) \indone{ [H_{I_n}, [H_{I_n}, \rho_S \otimes \rho_{E_n}]]} \\
    & \leq \norm{B^2}_\infty \indone{S^2\rho_S + \rho_S S^2 - 2 S \rho_S S}\\
    & \leq \norm{B^2}_\infty (2\norm{\rho_S}_1 \norm{S^2}_\infty + 2\norm{\rho_S}_1 \norm{S}_\infty^2) \\
    &\leq 4 \norm{B^2}_\infty \norm{S}_\infty^2
\end{align}
where $e_j(B)$ represent the ordered eigenvalues of the matrix $B$. We the von Neumann trace inequality in the second line, and use identity from Ref. \cite{watrous2018theory} P.33 (Eq. 1.175) in the third line. We could proceed to bound the remainder of the terms outside the sums in the same fashion, however, the takeaway here is that without using the limiting RI map $\widetilde{\mathcal{R}_n} (\tau)$, we obtain an error in the first non-trivial order in the interaction time $\tau$. To leading order, again with smallness assumptions, we have:
\begin{equation}
    \epsilon_\lambda \in O(\tau \norm{B^2}_\infty \norm{S}_\infty^2).
\end{equation}
Therefore, implementing an RI map without the limiting case of $\lambda^2 \tau \rightarrow 1$, would necessarily impose severe restrictions on the interaction time $\tau$, and consequentially on number of required interactions $m$ which would result in poor complexity bounds. For this reason, we focus only RI maps where our desired limit or replacement of variables holds.

%======================
% Appendix B

\section{Sufficiency Criteria} \label{sec:appendixB}

In this Section we provide some sufficiency criteria that allow for a map of the RI form to capture Lindblad dynamics. We show these criteria through a derivation and summarize them at the end. It must be stressed that these criteria are not necessary conditions, meaning they are not \textit{if and only if} statements. It may be entirely possible to find other kinds of interaction Hamiltonians that lead to a Lindblad form. Fully classifying these Hamiltonians, as well as investigating the other direction of deriving an RI map from a Lindblad equation is an interesting open problem. Broadly speaking, the jump operators $L_j$ cannot be entirely arbitrary and must be \textit{derivable} from an interaction Hamiltonian with a partial trace over the environment Hilbert space. We precisely show this by first defining our Lindbladian in the form from Section \ref{sec:errorbound} restated below.
\begin{equation*} 
    \liouv(\rho_S) = -i[H_0, \rho] - \frac{1}{2}\sum_{k=1}^m\Tr_{\mathscr{H}_E} [H_{I_k}, [H_{I_k}, \rho_S \otimes \rho_{E_k}]] .
\end{equation*}
This is the Lindbladian that can be implemented with RI maps in general. Now recall the fully general Lindblad equation with arbitrary jump operators $L_j$: 
\begin{equation*}
    \liouv(\rho_S) = -i[H_0, \rho_S] + \sum_{j=1} ^m \left[ L_j \rho_S L_j^\dagger - \frac{1}{2} \{L_j^\dagger L_j , \rho_S \} \right]. 
\end{equation*}
Our goal in this Section is to establish sufficiency criteria on the interaction Hamiltonians $H_{I_k}$ and the environment density matrix $\rho_{E_k}$ to recover the above form. Looking at the two, it is clear that the system Hamiltonian can be entirely arbitrary. As well, neither Lindbladian depends on the Hamiltonian of the environment from the RI map (although the convergence parameter $\nu$ scales with the norm of this Hamiltonian). In addition, if we show criteria sufficient to recover a single jump operator $L_j$ from an interaction Hamiltonian and bath element, then it will hold for all $m$ jumps. Thus, we want to find criteria for the following to hold:
\begin{align}
    L_j \rho_S L_j^\dagger - \frac{1}{2} \{L_j^\dagger L_j , \rho_S \} &= -\frac{1}{2}\Tr_{\mathscr{H}_E} [H_{I_k}, [H_{I_k}, \rho_S \otimes \rho_{E_k}]] \\
    & = \Tr_{\mathscr{H}_E} \Bigr ( H_{I_k} \rho_S \otimes \rho_{E_k} H_{I_k} - \frac{1}{2}\{H_{I_k} H_{I_k} ,\rho_S \otimes \rho_{E_k} \} \Bigr ).
\end{align}
Next, let us write the interaction Hamiltonian in the usual form $H_{I_k} = S_k \otimes B_k^\dagger + S_k^\dagger \otimes B_k$. The analysis is simpler if we consider a term wise comparison. First examining the left most term above:
\begin{align}
    \Tr_{\mathscr{H}_E} \Bigr ( H_{I_k} \rho_S \otimes \rho_{E_k} H_{I_k} \Bigr ) &= S_k^\dagger \rho_S S_k \Tr_{\mathscr{H}_E} \Bigr (B_k \rho_{E_k} B_k^\dagger \Bigr ) + S_k \rho_S S_k^\dagger \Tr_{\mathscr{H}_E} \Bigr (B_k^\dagger \rho_{E_k} B_k \Bigr ) \\
    & + S_k \rho_S S_k \Tr_{\mathscr{H}_E} \Bigr ((B_k^\dagger)^2 \rho_{E_k} \Bigr ) + S_k^\dagger \rho_S S_k^\dagger \Tr_{\mathscr{H}_E} \Bigr (B_k^2 \rho_{E_k} \Bigr ), 
\end{align}
now if we define 
\begin{equation} \label{eq:Lj}
    L_j = S_k \sqrt{z_k} \: \text{with} \:  z_k = \Tr_{\mathscr{H}_E} \Bigr (B_k^\dagger \rho_{E_k} B_k \Bigr ),
\end{equation}
 and 
 \begin{equation}\label{eq:Ljplus1}
     L_{j+1} = S_k^\dagger \sqrt{\Bar{z_k}} \: \text{with} \: \Bar{z_k} = \Tr_{\mathscr{H}_E} \Bigr (B_k \rho_{E_k} B_k^\dagger \Bigr ),
 \end{equation}
 and enforce that
\begin{equation} \label{eq:bath_criteria}
    \Tr_{\mathscr{H}_E} \Bigr (B_k^2 \rho_{E_k} \Bigr ) = \Tr_{\mathscr{H}_E} \Bigr ((B_k^\dagger)^2 \rho_{E_k} \Bigr ) = 0,
\end{equation}
then we obtain the following:
\begin{equation}
    \Tr_{\mathscr{H}_E} \Bigr ( H_{I_k} \rho_S \otimes \rho_{E_k} H_{I_k} \Bigr ) = L_j \rho_S L_j^\dagger + L_{j+1} \rho_S  L_{j+1}^\dagger,
\end{equation}
which is the Lindblad form for this term. Repeating this process for each $k$ leads to a Lindblad equation with $m \rightarrow 2m$ jump operators. Note the form of each jump operator is entirely general, with the constraint that its hermitian conjugate must be included as another jump operator. In the physical perspective, this is a consequence of the energy preserving nature of the system-bath interaction. The rates, or constants that multiply the jumps can be modulated by choosing the $B_k$ operators and the bath state $\rho_{E_k}$. Perhaps the condition that looks most restrictive here is that in Equation \ref{eq:bath_criteria}. However, in the case of our analysis, this holds for qubit creation and annihilation operators when the bath is a single qubit thermal state. More general, if we work with qubit operators, it is simple to show that this holds when $\rho_{E_k}$ has no coherences (i.e. only has support in the space $\text{diag}(\rho_{E_k})$) while $\text{diag}(B_k)=0$. Carrying these constraints through the second term: 
\begin{align}
    -\frac{1}{2} \Tr_{\mathscr{H}_E} \Bigr (\{H_{I_k} H_{I_k}, \rho_{E_k}\} \Bigr ) &= -\frac{1}{2} \Biggr (S_k S_k^\dagger \rho_S \Tr_{\mathscr{H}_E} \Bigr (B_k^\dagger B_k \rho_{E_k} \Bigr ) + S_k^\dagger S_k \rho_S \Tr_{\mathscr{H}_E} \Bigr (B_k B_k^\dagger \rho_{E_k} \Bigr ) \\
    & + \rho_S S_k^\dagger S_k  \Tr_{\mathscr{H}_E} \Bigr ( \rho_{E_k} B_k B_k^\dagger  \Bigr ) + \rho_S S_k S_k^\dagger  \Tr_{\mathscr{H}_E} \Bigr ( \rho_{E_k} B_k^\dagger B_k  \Bigr ) \Biggr ) \\
    & = -\frac{1}{2} \Biggr ( L_{j+1} L_{j+1}^\dagger \rho_S + L_j L_j^\dagger \rho_S +  \rho_S L_{j+1} L_{j+1}^\dagger  + \rho_S L_j  L_j^\dagger \Biggr ) \\
    & = -\frac{1}{2} \{L_j^\dagger L_j, \rho_S \} -\frac{1}{2} \{L_{j+1} L_{j+1}^\dagger, \rho_S \},  
\end{align} 
we have recovered the Lindblad form. When this process is repeated over all $m$ indices (for each corpuscle of the environment) the full Lindblad equation is recovered. Now this Section can be simply summarized: Given we have an interaction Hamiltonian written as $H_{I_k} = S_k \otimes B_k^\dagger + S_k^\dagger \otimes B_k$, and Equation \ref{eq:bath_criteria} is satisfied, then we have 
\begin{align}
    \liouv(\rho_S) &= -i[H_0, \rho] - \frac{1}{2}\sum_{k=1}^m\Tr_{\mathscr{H}_E} [H_{I_k}, [H_{I_k}, \rho_S \otimes \rho_{E_k}]] \\
    &= -i[H_0, \rho_S] + \sum_{j=1} ^{2m} L_j \rho_S L_j^\dagger - \frac{1}{2} \{L_j^\dagger L_j , \rho_S \}, 
\end{align}
with jump operators defined in Equations \ref{eq:Lj} and \ref{eq:Ljplus1}. Note that there are no constraints on the following operators: $H_0, \; H_{E_k}, \; S_k, \; \rho_S$, given that the usual properties of a Hamiltonian and density matrix are satisfied, which makes this approach quite general even when imposing our sufficiency conditions. We see in this section that a Lindblad equation naturally can be derived from any RI model, however, the converse appears to be an interesting open problem.

\section{Numerical Example} \label{sec:appendixC}

\begin{figure}[htbp!] 
    \centering\includegraphics[width=0.8\textwidth]{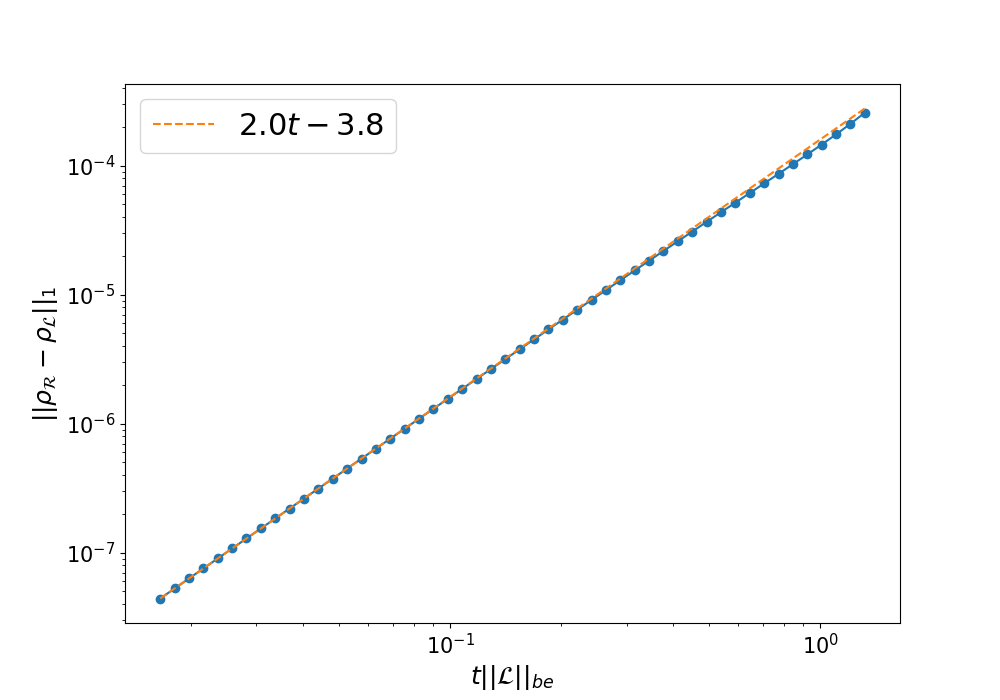}
    \caption{\textit{A plot of the error between the RI map and Lindbladian dynamics for a dissipative 4 site Heisenberg model described above. The error, for numerical computability purposes, is defined as the trace norm of the output density matrices from the respective channels. As expected, the error scales like $O(t^2 ||\liouv||^2)$ as proven in Theorem \ref{thm:RI_error}. The error begins to diverge in the neighbourhood of $t ||\liouv||_{be} =1$, where the numerical Taylor approximation to the Lindblad evolution no longer converges. The norm used on the Lindbladian, chosen for numerical convenience, is the block encoding norm as defined in both \cite{cleve2016efficient, li2022simulating}.
    Based on the discussion above, the following parameters were utilized to produce this plot: number of sites $N=4$ (implying $m=4$ for this model), bath frequency $\omega = 0.1$, inverse temperature $\beta =1$, magnetic field $b=0.5$, iteration parameter $\nu = 10$, block encoding norm $\norm{\liouv}_{be} = 16.5$ (calculated based on the model). These parameters are not particularly motivated, and can be arbitrarily chosen to produce the aforementioned scaling.}} \label{fig:error}
\end{figure}

Here we perform a numerical calculation of the error for the dynamics of a simple model to verify the result of Theorem \ref{thm:RI_error}. We wish to study the dissipative dynamics of a Heisenberg chain for short times and show that the dynamics agree with those of an RI model up to error that scales like $t^2$. Our Hamiltonian is as follows:
\begin{equation}
    H_0 = \sum_{k=1}^N  \Bigr (X_k X_{k+1} + Y_k Y_{k+1} + Z_k Z_{k+1} \Bigr ) + b Z_k, 
\end{equation}
where $b$ is a constant tuning the strength of the magnetic field relative to the interactions, and $\{X,Y,Z\}$ are the usual Pauli operators. For our RI map, we choose $H_{I_k} = V_k \otimes a^\dagger + V_k^\dagger \otimes a$, where $V_k = \openone_{2^{k-1}} \otimes a_k \otimes \openone_{2^{m-k}}$, which acts as a qubit annihilation operator (or spin lowering operator) at the $k$th site of the chain. For this example, we choose that this interaction occurs at each site, such that $m=N$.
Through initializing the environment to be single qubit thermal states, as was done in Section \ref{sec:RIsimulation}, this yields a Lindblad equation with jump operators of the form $L_j = V_k \sqrt{z_k}$ and $L_{j+1} = V_k^\dagger \sqrt{\bar{z_k}}$ where using Equations \ref{eq:Lj} and \ref{eq:Ljplus1}, $z = \frac{1}{1+ e^{-\beta}}$ and $\bar{z} = \frac{e^{-\beta}}{1+ e^{-\beta}}$. For simplicity, we set each element of the environment to the same inverse temperature $\beta$, and let them evolve under the same frequency $\omega$, such that their Hamiltonian is $H_E = \omega Z$. \\
For our numerical methods, we directly build the RI map from Equation \ref{eq:full_RI}, and use it to evolve a density matrix $\rho_\mathcal{R}$ for a time $t$. For the Lindblad dynamics, rather than vectorizing the density matrix equation, we simply perform a Taylor expansion of the exact dynamics to fourth order $e^{t \liouv} \approx \openone + t \liouv + \frac{t^2}{2} \liouv^2 + \frac{t^3}{6} \liouv^3 + \frac{t^4}{24} \liouv^4$, which is more than what is required for our purposes, considering that the RI map is a quadratic approximation in time. Here, an exponent on $\liouv$ means repeated applications of the map. We use this Taylor expansion to evolve another density matrix $\rho_\liouv$ for a time $t$. We then plot the 1-norm between these two density matrices in Figure \ref{fig:error}.

As proved in Theorem \ref{thm:RI_error}, the numerical calculations confirm that the error has the scaling $\epsilon \in O(t^2)$, confirming the aforementioned intuition that by iterating the RI map $\nu$ times, the error can be made arbitrarily small.

\end{document}